\documentclass[12pt]{article}
 \usepackage{mathtools}
 \usepackage{ mathrsfs }

\usepackage{bm}
\usepackage{graphics}
\usepackage{amsfonts}

\usepackage{amsthm}
\usepackage{amsmath}
\usepackage{latexsym}
\usepackage[all]{xy}
\usepackage{fancyhdr}
\usepackage{color}
\usepackage{cite}

\usepackage{amssymb}
\usepackage{amscd}
\usepackage{graphics}
\usepackage{amsfonts}
\usepackage{amsmath}
\usepackage{latexsym}
\usepackage[all]{xy}
 
\usepackage{empheq}

\usepackage{latexsym,graphics,color}
\def\bes{\begin{subequations}}
\def\ees{\end{subequations}}
\def\ba{\begin{align}}
\def\ea{\end{align}}

\def\bs{\boldsymbol}

\def\be{\begin{equation}}
\def\ee{\end{equation}}
\def\D{\mathcal D}

\def\K{\mathcal K}
\def\H{\mathcal {H}}
\def\E{\mathcal E}

\def\bc{\mathbb C}
\def\br{\mathbb R}

\def\F{\mathcal F}

\def\s1{\sigma^1}
\def\s2{\sigma^2}
\def\s3{\sigma^3}

\def\L{\mathcal L}

\def\si{\sigma}

\def\d{\partial}
\def\jp{\frac{1}{2}}
\def\ri{{\mathrm i}}

\def\Ad{{\rm Ad}}
\def\ad{{\rm ad}}

\definecolor{lila}{rgb}{1,0.2,0.9}
\definecolor{brown}{rgb}{0.5,0.3,0.3}
\definecolor{turquoise}{rgb}{0.2,0.9,0.7}
\definecolor{Orange}{rgb}{0.93,0.44,0}           
\definecolor{GrayBlue}{rgb}{0.35,0.4,0.62}       
\definecolor{SeafoamGreen}{rgb}{0.54,0.71,0.50}  

\definecolor{darkorange}{cmyk}{.20,.50,.80,0}
\definecolor{lightorange}{cmyk}{.07,.37,.65,0}
\definecolor{darkpeagreen}{cmyk}{.50,.30,.50,0}
\definecolor{lightpeagreen}{cmyk}{.22,.20,.40,0}

\newtheorem{thm}{Veta}[section]

\theoremstyle{definition}

\theoremstyle{definition}
\newtheorem{rem}[thm]{Remark}

\theoremstyle{definition}
\def\G{{\cal G}}                        %
\def\ri{{\mathrm{i}}}                   %
\def\1{{\mbox{\boldmath $1$}}}          %

\def\lm{\lambda}                        %
\def\jp{\frac{1}{2}}                    %
\def\om{\omega}                         %
\def\al{\alpha}                         %
\def\ga{\gamma}                         %

\definecolor{spec}{rgb}{0.0, 0.26, 0.15}

\def\bpm{\begin{pmatrix}}
\def\epm{\end{pmatrix}}

\DeclareMathSymbol{\Rho}{\mathalpha}{operators}{"50}

\begin{document}

\begin{flushright}
{}~
  
\end{flushright}

\vspace{1cm}
\begin{center}
{\large \bf  On strong integrability of the dressing cosets}

\vspace{1cm}

{\small
{\bf Ctirad Klim\v c\'ik}
\\
Aix Marseille Universit\'e, CNRS, Centrale Marseille\\ I2M, UMR 7373\\ 13453 Marseille, France}
\end{center}

\vspace{0.5 cm}

\centerline{\bf Abstract}
\vspace{0.5 cm}
\noindent  We formulate sufficient conditions for the strong integrability of dressing cosets. We provide several sigma-model backgrounds solving those conditions, some of them are new and some of them   were not so far formulated as the dressing cosets.   The new models are based on the Drinfeld doubles having the structure of higher order jet bundles of quadratic Lie groups.

  \vspace{2pc}
  
  \section{Introduction}
  The dressing cosets  are particular nonlinear $\sigma$-models in two dimensions which were originally  introduced in \cite{KS96b} in a successfull attempt to generalize the so-called Poisson-Lie T-duality \cite{KS95}. The first order Hamiltonian dynamics of the
  dressing cosets can be described  in terms of the dynamical systems referred to as degenerate
 $\hat\E$-models which themselves  generalize the so called non-degenerate $\E$-models.  This generalization is substantial, nevertheless it received a relatively little attention in the literature so far although there was  some older activity in the field \cite{BTW,CM06,HS,HT,SST15,S98,S99}, just like a more recent one \cite{SV,K19,K21,DHPT19,DHPT20,HLR,R21}.
  
  We believe that the dressing cosets as well as the degenerate $\hat\E$-models are going to play an increasing role in  future  in all those  places where their simpler non-degenerate counterparts already play role, like in the theory of integrable $\sigma$-models \cite{K02,K09,K14,S14,M20,B,Mad93,C81,Fa96,KMY,DHKM,DMV13,DDST18,OV}.   In particular, it is well-known 
 that many integrable $\sigma$-models on group manifolds can be  in fact interpreted in terms of the non-degenerate $\E$-models \cite{K15,K16,K17,K20,HL,LV20}. However,  in the case of the integrable models living on the coset manifold an unified  interpretation in terms of the dressing cosets is  so far missing. Among other things, we fill this gap at least partially in the present paper.
 
 It is important to distinguish between a weak and a strong integrability of a non-linear $\sigma$-model. The weak integrability means that the equation of motion of the model can be written
 in the so called Lax form  with spectral parameter $z$, that is in the
 form
 \be \frac{dL(z)}{d\tau}=[L(z),M(z)],\ee
 where $L(z)$ and $M(z)$ are $z$-families of  matrix-valued functions on the phase space of the $\sigma$-model. The strong integrability then means that every two spectral invariants  of the Lax matrix $L(z)$ Poisson commute, for whatever values of the spectral parameters $z$ and $w$.  As we shall see later on, this means that the matrix Poisson bracket of the Lax matrix $L(z)$ with $L(w)$  is governed by the so called $r$-matrix $r(z,w)$, which is an important characteristics of the strongly integrable model.
 
 Sufficient conditions for the weak integrability of the non-degenerate $\E$-models were first formulated in \cite{S17} and solved in many cases in \cite{LV20}.  They were later extended in \cite{K21} to ensure also the strong
integrability. In this paper, we formulate the sufficient conditions for the strong integrability of the degenerate $\hat\E$-models and we solve them for several classes
 of the dressing cosets. Some of those classes constitute a reinterpretation of the known symmetric spaces $\sigma$-models in terms of the dressing cosets while the others are completely new and are based on  high order jet bundles of quadratic Lie groups interpreted as the Drinfeld doubles.
 In particular, the jet bundle $J^{2n+1}G$ of a compact semi-simple Lie group $G$ gives rise to the strongly integrable dressing coset describing a suitable interaction of $n$ fields with values in the Lie algebra $\G$ of the group $G$.
 
 The plan of the paper is as follows: In Section 2, we review the concept of the non-degenerate $\E$-model as well as the sufficient conditions of its integrability (the weak as well as the strong ones). In Section 3.1, we review in much technical detail the concepts
 of the degenerate $\hat\E$-models as well as that of the dressing cosets while in Section 3.2 we start with the presentation of the original results, namely, we formulate the sufficient conditions for the strong  integrability of the dressing cosets. In Section 4, we interpret a non-deformed, a $\lm$-deformed and an $\eta$-deformed symmetric space $\sigma$-models as the dressing cosets and we show that the well-known strong integrability of those three theories can be understood in an unified way within our degenerate $\hat\E$-model formalism.  In Section 5, we solve our sufficient conditions of the strong integrability   for 
 a new family of the dressing cosets which is  parametrized by  positive integers. Thus, for given $n$, the model describes a suitable integrable interaction of  $n$ $\G$-valued fields. Only for $n=1$ the resulting $\sigma$-model was known previously,  it is in fact  the pseudo-chiral  model of Zakharov and Mikhailov \cite{ZM}.
 We determine also the so called twist function of the generalized pseudo-chiral integrable model for every $n$.  In Section 6, we provide   conclusions and an outlook.

   \setcounter{equation}{0}
  \section{Reminder: non-degenerate $\E$-models}
  \setcounter{equation}{0}
 \subsection{The first and  the second order formulations}
 
  In this paper, by a Drinfeld double $D$ we shall understand a connected even-dimensional
 Lie group   equipped with a bi-invariant pseudo-Riemannian metric of maximally-Lorentzian (split) signature.
 An $\E$-model, introduced in \cite{KS95,KS96a,K15}, is a first-order Hamiltonian dynamical system $(\omega,H_\E)$ living on the loop group $LD$ of the Drinfeld double. The symplectic form $\omega$ and  the Hamiltonian $H_\E$ on $LD$ are respectively given by the
 formulas
   \be   \omega =-\jp\oint   \bigl(l^{-1}\delta l\stackrel{\wedge}{,}(l^{-1}\delta l)'\bigr)_{\D}\label{symf},\ee
 \be H_\E  =\jp\oint \bigl(l'l^{-1},\E \ \! l'l^{-1}\bigr)_{\D}.\label{Ham}\ee
Here $(.,.)_\D$ is the non-degenerate ad-invariant symmetric bilinear form $(.,.)_\D$ defined on the Lie algebra $\D$ of the Drinfeld double (it is given by the pseudo-Riemannian metric at the group origin). The integration in Eqs. \eqref{symf}, \eqref{Ham} is over the loop parameter $\sigma$, while the derivative with respect to the
 loop parameter is denoted by the apostrophe. The symbol $\delta$ stands for the de Rham exterior differential on the loop group $LD$, the closed string configuration $l$ parametrizes $LD$ and, finally,
 $\E:\D\to\D$ is an $\br$-linear operator squaring to identity, symmetric with respect to the bilinear form $(.,.)_\D$ and such that the bilinear form $(.,\E.)_\D$
 on $\D$ is strictly positive definite.
 
 \medskip
 
  \medskip
 
 It turns out, that the left action  of the group $LD$ on itself  
 is generated by the moment map $j=l'l^{-1}$.
 The components of the current  $j$ then verify the Poisson current algebra
 \be \{(j(\sigma_1),T_1)_\D,(j(\sigma_1),T_2)_\D\}=(j(\sigma_1),[T_1,T_2])_\D\delta(\sigma_1-\sigma_2)+(T_1,T_2)_\D\delta'(\sigma_1-\sigma_2)\label{spc}\ee
 which plays a crucial role in the dynamics of the $\E$-models. Indeed, the Poisson brackets $\eqref{spc}$ as well as the explicit form \eqref{Ham} of the Hamiltonian give the following
 first order equations of motion of the $\E$-model
 \be \frac{\d j}{\d \tau}=\{j,H_\E\}=(\E j)'+[\E j,j], \label{strfe}\ee

Consider the $\E$-model on the Drinfeld double $D$ and let $K\subset D$ be a  half-dimensional subgroup   such that the Lie subalgebra $\K$ is isotropic with respect to the bilinear form $(.,.)_\D$. Then it was shown in \cite{KS97} that there is a two-dimensional non-linear $\sigma$-model such that its first order dynamics can be expressed in terms of the $\E$-model; in particular
its first order Hamiltonian equations of motion are given by \eqref{strfe}. The action of this $\sigma$-model reads
$$S_{\E}(l)=\frac{1}{4}\int \delta ^{-1}\oint \biggl(\delta ll^{-1},[\partial_\sigma ll^{-1},\delta ll^{-1}]_\D\biggr)_\D +$$\be +\frac{1}{4} \int d\tau\oint \biggl(W^+_l\partial_+ ll^{-1}, \partial_- ll^{-1}\biggr)_\D-\frac{1}{4} \int d\tau\oint \biggl( \partial_+ ll^{-1},W^-_l \partial_- ll^{-1}\biggr)_\D. \label{2nd} \ee
 Here $l(\tau,\sigma)\in D$ is a field configuration, $\d_\pm=\d_\tau\pm\d_\si$,
 $\delta^{-1}$ is a (symbolic) inverse of the de Rham differential and $W^\pm_l:\D\to\D$ are the projectors  fully characterized by their respective   kernels and images
\be \label{projectors}  \mathrm{Ker}(W_l^\pm)=Ad_l(\K),  \quad
  \mathrm{Im}(W_l^\pm)=(1\pm\E)\D.\ee
 It may seem, that the $\si$-model \eqref{2nd} lives  on the target $D$ but, actually, it lives on the space of cosets $D/K$ because
 it  enjoys  the gauge symmetry
 \be \label{gs}
  l(\tau,\sigma)\to l(\tau,\sigma)k(\tau,\sigma), \quad k(\tau,\si)\in K.\ee
  In some cases, the fibration $D\to D/K$ admits a  global section $p:D/K\to D$, in which case we can fix the gauge
  $l=p$ in the action \eqref{2nd}.
  
  Note also that the equations of motion of the $\sigma$-model \eqref{2nd} can be written as
 \be \d_+(W^-_l\d_-ll^{-1})-\d_-(W^+_l\d_+ll^{-1})
 -\left[W^+_l\d_+ll^{-1},W^-_l\d_-ll^{-1}\right]_\D=0.\label{qeq}\ee

  \subsection{Integrability of the non-degenerate $\E$-models}

 Suppose that there is the $\E$-model based on the  Drinfeld double $D$ and  there is also a quadratic  Lie algebra $\G$ together with an one-parametric family
 of linear operators
 $O(z):\D\to\G$ 
verifying the following conditions
 \be [O(z)x_+, O(z)x_-]_\G=O(z)[x_+,x_-]_\D, \quad x_\pm\in\D,\quad  \E x_\pm=\pm x_\pm.\label{sufc}\ee

  \medskip
  
 It was shown in \cite{S17,LV20}  that Eq. \eqref{sufc} is the sufficient condition for the existence of the Lax pair $L(z),M(z)$ of this $\E$-model.
 The operators $L(z),M(z)$ act on the loop Lie algebra
  $L\G$ and they are  given by the formulas
    \be L(z)  =\d_\si - \ad^\G_{O(z)j},\label{Laxfh}\ee 
    \be M(z)=- \ad^\G_{O(z)\E j}.\label{M}\ee
The condition  \eqref{sufc} guarantees that the field equations \eqref{strfe} of the $\E$-model  can be represented in the Lax form with  spectral parameter \cite{La}
        \be  \{L(z),H_\E\}= \frac{dL(z)}{dt} =[L(z),M(z)].\label{Laxcondition}\ee
        
The Lax form \eqref{Laxcondition} of the field equations guarantees the so called weak Lax integrability, which means that the
    spectral invariants (typically the traces of the powers) of the operator $L(z)$ are the integrals of motion. 
  If, however,  those integrals of motions Poisson commute among each other, we say that the $\E$-model is strongly Lax integrable \cite{BV,M,S}.
  
  The sufficient conditions for the strong Lax integrability of $\E$-models were formulated in \cite{K21}. It is thus required that it exists a non-degenerate invariant symmetric bilinear form $(.,.)_\G$ on $\G$ and there is a two-parametric family of operators $\hat r(z,w):\G\to\G$ such that it holds
  \be [O^\dagger(z)x,O^\dagger(w)y]_\D+O^\dagger(z)[x,\hat r(z,w)y]_\G+O^\dagger(w)[\hat r(w,z)x,y]_\G=0,\quad \forall x,y\in\G,\label{1d}\ee
  \be (O^\dagger(z)x,O^\dagger(w)y)_\D+(x,\hat r(z,w)y)_\G+(\hat r(w,z)x,y)_\G=0,\quad \forall x,y\in\G.\label{2d}\ee
   Here   $O(z)^\dagger:\G\to\D$
 is  the adjoint  of the operator $O(z)$; it is defined by the relation
  \be (O(z)x,y)_\G=(x,O^\dagger(z)y)_\D, \quad  \forall x\in\D, y\in\G.\label{ado}\ee
  
 To see, why \eqref{1d} and \eqref{2d} are 
 the sufficient conditions of the strong Lax integrability, we  use them as well as  Eq.\eqref{spc} to calculate the Poisson brackets of  the matrix elements of the Lax operator $L(z)$ 
 $$\{(y',L(z)(\si_1)x')_\G,(y'',L(w)(\si_2)x'')_\G\}=\{(O(z)j(\si_1),x)_\G,(O(w)j(\si_2),y)_\G\}=$$$$ =(j(\si_1),[O^\dagger(z)x,O^\dagger(w)y]_\D)_\D\delta(\si_1-\si_2)  +(O^\dagger(z)x,O^\dagger(w)y)_\D\d_{\si_1}\delta(\si_1-\si_2)= $$
  $$= -\Bigl((O(z)j(\si_1),[x,\hat r(z,w)y]_\G)_\G+(O(w)j(\si_2),[\hat r(w,z)x,y]_\G)_\G\Bigr)\delta(\si_1-\si_2)$$\be -\Bigl((x,\hat r(z,w)y)_\G+(\hat r(w,z)x,y)_\G\Bigr)\d_{\sigma_1}\delta(\si_1-\si_2) .\label{por}\ee
 where  we have set
 \be x=[y',x']_\G,\quad y=[y'',x'']_\G,\quad  x',x'',y',y''\in\G.\ee
 The relation \eqref{por} can be rewritten in the operator form as
 \be \{L(z)\otimes {\rm Id},{\rm Id}\otimes L(w)\}=[r(z,w),L(z)\otimes {\rm Id}]-[r^p(w,z),{\rm Id}\otimes L(w)],\label{si}\ee
 where the operator $r(z,w)$ acts on the tensor product $L\G\otimes L\G$ and it is defined as   \be r(z,w)=C_{AB}\ad_{\hat r(z,w)T^A}\otimes\ad_{T^B}\delta(\si_1-\si_2).\label{ans1}\ee
  Note that $C_{AB}$ is the inverse matrix of $C^{AB}$ defined by 
     \be C^{AB}:=(T^A,T^B)_\G,\ee
where $T^A$ is some basis of the Lie algebra $\G$ on the choice of which actually the $r$-matrix \eqref{ans1} does not depend.
The notation $r^p$ means  
  \be r^p=\sum_\alpha B_\alpha\otimes A_\alpha,\ee
 if $r$ has the form
  \be r=\sum_\alpha A_\alpha\otimes B_\alpha \ee
  for some family of linear operators $A_\alpha,B_\alpha$ acting on $L\G$.  The spectral parameter $z$ may even take complex values
  in which case the crucial map $O(z)$ is considered as the map from
  $\D$ to $\G^\bc$.
  
  \medskip
  
  By the general theory of the Lax
  integrable system exposed e.g. in the book \cite{BBT}, the validity of the identity \eqref{si} precisely guarantees the strong integrability of the dynamical system possessing the Lax operator $L(z)$. 
        
        \medskip

  \section{Degenerate $\hat\E$-models}
   \setcounter{equation}{0}
  \subsection{Reminder: dressing cosets}

  Consider again the $\br$-linear involution $\E:\D\to\D$ symmetric with respect to the bilinear form $(.,.)_\D$ and such that the bilinear form $(.,\E.)_\D$
 on $\D$ is strictly positive definite.
 Furthermore, we suppose
 
 \smallskip
 
 \noindent 1) There is an isotropic subgroup $F\subset D$ such that the involution
 $\E$ commutes with the adjoint action of $F$ on $\D$;
 
 \smallskip
 
 \noindent 2) The
 restriction of the bilinear form $(.,\E.)_\D$ to the Lie subalgebra $\F$ is strictly positive definite.
 
 \medskip
 
In \cite{K19},   the so called {\it degenerate $\hat\E$-model}  was  associated 
 to the data just described 
 as a slight reformulation of the so called   {\it dressing coset} introduced in \cite{KS96a} (in what follows, we use the both terms interchangeably).
 Thus the degenerate 
 $\hat\E$-model is  the first-order Hamiltonian dynamical system  
 obtained by an appropriate symplectic reduction of the dynamical system
 $(\omega,H_{\hat\E})$, where the symplectic
 form $\omega$ is the standard Kirillov-Drinfeld one
     \be   \omega =-\jp\oint   \bigl(l^{-1}\delta l\stackrel{\wedge}{,}(l^{-1}\delta l)'\bigr)_{\D}\label{symfbis},\ee
     while the Hamiltonian is given by
 \be H_{\hat\E}(l)  =\jp\oint \bigl(j,\hat\E j\bigr)_{\D}, \quad  j:=l'l^{-1}.\label{Hambis}\ee
 Note that the dynamical system $(\omega,H_{\hat\E})$
that we are going to reduce symplectically {\it is not} the
non-degenerate $\E$-model in the sense of the definition given in Section 2,  inspite of the fact that the symplectic form \eqref{symfbis} is the same as \eqref{symf}. This is because
the linear operator $\hat\E$ appearing in the Hamiltonian \eqref{Hambis} is not the same thing as $\E$ although it is obtained from $\E$ in an appropriate way. Actually, $\hat\E$  has a nontrivial kernel and, therefore, it cannot be involutive. To define $\hat\E$, we first notice that the condition 2) above means that 
the intersection of the Lie subalgebra $\F\subset\D$ with the linear space $\E\F$ is trivial (it contains just $0\in\D$), which means that the Drinfeld double Lie algebra $\D$ can be represented as the direct sum of four $\Ad_F$-invariant vector spaces as follows
\be \D=V_+\oplus V_-\oplus \F\oplus \E\F,\label{dirsum}\ee
where the subspaces $V_\pm$ are defined by the conditions
\be V_+\oplus V_-=(\F\oplus\E\F)^\perp,\quad \E x_\pm=\pm x_\pm,\quad x_\pm\in V_\pm.\label{vpm}\ee
Note that the notation $\perp$ means the perpendicularity with respect to the bilinear form $(.,.)_\D$.

Following the decomposition \eqref{dirsum}, every element $y\in\D$ can be unambiguosly written as
\be y=y_++y_-+y^\F+y^{\E\F},\ee
 and we can now define the operator $\hat\E:\D\to\D$  by the formula
 \be \hat\E y:=\E (y_++y_-+ y^{\E\F}).\ee
 Occasionally, we shall also  use the notation
 \be y^\perp:=y_++y_-, \label{ppm}\ee
 in particular we write
 \be y=y^\perp+y^\F+y^{\E\F}\ee
 and
 \be \hat\E y=\E(y^\perp+y^{\E\F}).\ee
 
 It turns out that the dynamical system $(\om,H_{\hat\E})$ can be indeed symplectically reduced 
 with respect to the left action of the loop group $LF$ on the loop group $LD$. To see it, we have to prove that both the symplectic form \eqref{symfbis} and the Hamiltonian \eqref{Hambis} are invariant with respect to the action of $LF$ on the unreduced phase space  $LD$.  
 
 First we show the $LF$-invariance of the Hamiltonian $H_{\hat\E}$. Let $f\in LF$.
 Then we have
$$ H_{\hat\E}(fl)   =\jp\oint \bigl((fl)'(fl)^{-1},\hat\E \ \! (fl)'(fl)^{-1}\bigr)_{\D}=$$$$=\jp\oint \bigl(f'f^{-1}+\Ad_f((l'l^{-1})^\F + (l'l^{-1})^\perp +( l'l^{-1})^{\E\F}),\E\Ad_f((l'l^{-1})^\perp +( l'l^{-1})^{\E\F})\bigr)_\D=$$
\be =\jp\oint \bigl( (l'l^{-1})^\F +(l'l^{-1})^\perp +( l'l^{-1})^{\E\F},\E((l'l^{-1})^\perp +( l'l^{-1})^{\E\F})\bigr)_\D=H_{\hat\E}(l) \label{Hamtris}\ee
Now we show the $LF$-invariance of the symplectic form $\omega$. Let $\phi(\si)\in L\F$ and consider the vector field $v_\phi$ acting an a function $\Psi$ on $LD$ as
\be (v_\phi\Psi)(l):=\left(\frac{d}{ds}\right)_{s=0}\Psi(e^{s\phi}l).\ee
Since $v_\phi$ is the right-invariant vector field on the loop group $LD$, its contraction with the right-invariant Maurer-Cartan form $\delta ll^{-1}$ on $LD$ is
\be \langle \iota_{v_\phi},\delta ll^{-1}\rangle=\phi.\label{cont}\ee
Then we deduce from Eqs. \eqref{symfbis} and \eqref{cont}
\be \iota_{v_\phi}\omega=\delta(l'l^{-1},\phi)_\D=\delta(j,\phi)_\D=\delta(j^{\E\F},\phi)_\D,\label{ivo}\ee
which means that the quantity $-(j^{\E\F},\phi)_\D$ is the moment map for the left action of the element $\phi$ on $LD$.

Since the form $\omega$ is closed, its Lie derivative with respect to the vector field $v_\phi$ vanishes. Indeed, we have
\be \L_{v_\phi}\omega=(\delta\iota_{v_\phi}+\iota_{v_\phi}\delta)\omega=\delta(\delta(j^{\E\F},\phi)_\D)=0.\ee
 
 \medskip
 
 Now we can perform the symplectic
 reduction of the dynamical system
 $(\om,H_{\hat\E})$ by setting
 \be j^{\E\F}=0.\label{rcon}\ee
 The reduced dynamical system is denoted as
 $(\hat\omega,\hat H_{\hat\E})$
and it is
 referred to as the degenerate $\hat\E$-model. The reduced phase space
 $\hat P$ can be identified with the space of the 
left cosets $LF\backslash LD^c$, where $LD^c$ is the space of all elements of the loop group $LD$ verifying the $LF$-invariant constraint  \eqref{rcon}. The reduced Hamiltonian $\hat H_{\hat\E}$ is the
restriction of the unreduced Hamiltonian $ H_{\hat\E}$ from the
space $LF\backslash LD$ to the space $LF\backslash LD^c$.

\medskip

\begin{rem}{\small
Note that even if the involution $\E$ satisfies the conditions 1) and 2) above, the Hamiltonian \eqref{Ham} of the non-degenerate $\E$-model cannot be reduced because it is not
invariant  with respect to the  left action of the loop group $LF$ on the loop group $LD$. For this reason, the degenerate $\hat\E$-model cannot be interpreted as the symplectic reduction of the non-degenerate one.}
\end{rem}

The equations of motion of the
 unreduced dynamical system $(\omega,H_{\hat\E})$ 
obviously read
 \be \frac{\d j}{d \tau}=\{j,H_{\hat\E}\}=(\hat\E j)'+[\hat\E j,j]_\D, \label{strdr}\ee
 or, in the decomposed form following the decomposition \eqref{dirsum}, as
 \begin{subequations} \label{feun}\begin{align}\frac{\d j_+}{\d \tau}&=j_+'+2[j_+,j_-]_+ +[\E j^{\E\F},j_+]-[j^\F,j_+]-[j^{\E\F},j_+-j_-]_+,\\ \frac{\d j_-}{\d \tau}&=-j_-'+2[j_+,j_-]_- +[\E j^{\E\F},j_-]+[j^\F,j_-]-[j^{\E\F},j_+-j_-]_-,\\ \frac{\d j^\F}{\d \tau}&=\E(j^{\E\F})'+2[j_+,j_-]^\F +[\E j^{\E\F},j^\F] -[j^{\E\F},j_+-j_-]^\F,\\ \frac{\d j^{\E\F}}{\d \tau}&=  [\E j^{\E\F},j^{\E\F}]-[j^{\E\F},j_+-j_-]^{\E\F}.\end{align}\end{subequations}
Imposing the reduction constraint $j^{\E\F}=0$, we obtain the equations of motions of the
  degenerate $\hat\E$-model  
   \begin{subequations} \label{fere}\begin{align}\frac{\d j_+}{\d \tau}&=j_+'+2[j_+,j_-]_+ -[j^\F,j_+] ,\\ \frac{\d j_-}{\d \tau}&=-j_-'+2[j_+,j_-]_-  +[j^\F,j_-] ,\\ \frac{\d j^\F}{\d \tau}&= 2[j_+,j_-]^\F .\end{align}\end{subequations}
   It is perhaps worth pointing out that the
   quantities $j_\pm$ and $j^\F$ appearing in the equation \eqref{fere} {\it are not} well defined on the reduced phase space $LF\backslash LD^c$ because  we find from the identity $j=l'l^{-1}$ that they transform  under the action of the loop group $LF$ as
\be j_\pm \to \Ad_fj_\pm, \quad j^\F\to \Ad_f j^\F+\d_\sigma ff^{-1},\quad f\in LF.\label{ctrtris}\ee
However, the equations \eqref{fere} themselves {\it are} invariant with respect to the transformation \eqref{ctrtris}, they  can be therefore indeed interpreted as the the evolution equations on the reduced phase space.
   
   \medskip

\begin{rem}{\small
   We note that the same set of the equations \eqref{fere} can be obtained by identifying $A_\sigma=j^\F$  and by fixing the gauge $A_\tau=0$  in the following set of the $F$-gauge invariant equations (that is the gauge transformation depends also on $\tau$)
     \begin{subequations} \label{gaufix}\begin{align}\d_- j_+-[A_-, j_+]&=2[j_+,j_-]_+,\\ \d_+ j_--[A_+, j_-]&=2[j_+,j_-]_- ,\\ \jp \d_-A_+-\jp\d_+A_-+\jp[A_+,A_-]&= 2[j_+,j_-]^\F ,\end{align}\end{subequations}
   where
   \be \d_\pm :=\d_\tau\pm\d_\si,\quad A_\pm :=A_\tau\pm A_\sigma.\ee
   The transformation \eqref{ctrtris} can be then interpreted as the residual time-independent gauge transformation preserving the  gauge $A_\tau =0$.}
   \end{rem}

\medskip
 
Consider the degenerate $\hat\E$-model on the Drinfeld double $D$ and let $K\subset D$ be a  half-dimensional subgroup   such that the Lie subalgebra $\K\subset\D$ is isotropic with respect to the bilinear form $(.,.)_\D$. Then there is a two-dimensional non-linear $\sigma$-model such that its first order dynamics can be expressed in terms of this degenerate $\hat\E$-model; in particular
its first order Hamiltonian equations of motion are given by \eqref{fere}. The action of this $\sigma$-model reads \cite{K21}
$$S_{\textrm{WZW}}(l)=\frac{1}{4}\int \delta ^{-1}\oint \biggl(\delta ll^{-1},[\partial_\sigma ll^{-1},\delta ll^{-1}]_\D\biggr)_\D +$$\be +\frac{1}{4} \int d\tau\oint \biggl(W^+_l\partial_+ ll^{-1}, \partial_- ll^{-1}\biggr)_\D-\frac{1}{4} \int d\tau\oint \biggl( \partial_+ ll^{-1},W^-_l \partial_- ll^{-1}\biggr)_\D. \label{2ndd} \ee
 Here   $W^\pm_l:\D\to\D$ are the projectors  fully characterized by their respective  kernels and images
\be \label{projectorsbis}  \mathrm{Ker}(W_l^\pm)=Ad_l(\K),  \quad
  \mathrm{Im}(W_l^\pm)=V_\pm\oplus\F.\ee
 It may seem, that  the $\si$-model \eqref{2ndd} lives  on the target $D$ but, actually, it lives on the space of double cosets $F\backslash D/K$ because
 it  enjoys  the gauge symmetries
 \be \label{gsbis}
  l(\tau,\sigma)\to f(\tau,\sigma)l(\tau,\sigma)k(\tau,\sigma), \qquad f(\tau,\si)\in F,\quad  k(\tau,\si)\in K.\ee

\subsection{Integrable  dressing cosets}
 
 Suppose that there is the degenerate $\hat\E$-model based on the Drinfeld double $D$, on the involution $\E$ and on the isotropic subalgebra $\F\subset\D$. Moreover, we suppose that there is a quadratic Lie algebra $\G$ which  also  possesses $\F$ as its   subalgebra and that there is an one-parametric  family
 of linear operators
 $\hat O(z):V_+\oplus V_-\to\G$ 
 which intertwin the adjoint action of $\F$ on $V_+\oplus V_-$ and on $\G$ and verify the following condition
 
\be  [\hat O(z)x_+, \hat O(z)x_-]_\G=[x_+,x_-]_\D^\F+\hat  O(z)[x_+,x_-]_\D^\perp, \quad  x_\pm\in V_\pm. \label{sufdr}\ee

\medskip

\noindent Here the element $[x_+,x_-]_\D^\F$ of $\F$ is viewed as the element of $\G$.

 \medskip

  It turns out that the condition \eqref{sufdr}  is sufficient  for the weak Lax integrability of the degenerate  $\hat\E$-model. Indeed, the Lax pair $L(z),M(z)$ 
  of the operators acting on the loop Lie algebra
  $L\G$ is given by the formulas
    \be L(z)  =\d_\si - \ad^\G_{j^\F +\hat O(z)j^\perp}\label{Laxfhbis}\ee
    \be M(z)=- \ad^\G_{\hat O(z)\E j^\perp }\label{Mbis}\ee
as it is straightforward to find out from  \eqref{sufdr} that  the dressing coset field equations \eqref{fere} of the
 degenerate $\hat\E$-model  can be indeed represented in the Lax form with the  spectral parameter
        \be  \frac{dL(z)}{dt} =[L(z),M(z)].\label{Laxconditionbis}\ee
        
        \medskip
        
      To guarantee the
        strong Lax integrability of the degenerate $\hat\E$-model, we first define
         \be O(z)(j^\F+j^\perp+j^{\E\F}):=j^\F+\hat O(z)j^\perp\ee
         and we supplement the condition \eqref{sufdr}  with the condition 
    that it exists a family of operators $\hat r(z,w):\G\to\G$ such that it holds 
  \be [ O^\dagger(z)x, O^\dagger(w)y]_\D+ O^\dagger(z)[x,\hat r(z,w)y]_\G+ O^\dagger(w)[\hat r(w,z)x,y]_\G\in \F,\quad  \forall x,y\in \G,\label{1e}
    \ee
  \be (O^\dagger(z)x, O^\dagger(w)y)_\D+(x,\hat r(z,w)y)_\G+(\hat r(w,z)x,y)_\G=0,\quad  \forall x,y\in \G.\label{2e}\ee
   Here $O^\dagger(z):\G\to\D$
 is  the adjoint  of the operator $O(z)$  defined by the relation
  \be ( O(z)x,y)_\G=(x,O^\dagger(z)y)_\D, \quad  \forall x\in\D, y\in\G.\label{adobis}\ee

        \medskip
        
     Let us now see why the conditions \eqref{1e} and \eqref{2e} guarantee the strong integrability. First we remark that the operators $\hat O(z)$ were supposed to 
     intertwin the adjoint action of $\F$ on $V_+\oplus V_-$ and on $\G$ 
 which implies that 
    the Lax operator $L(z)$   transforms in the adjoint way upon the action of the loop group $LF$. Indeed, using Eq. \eqref{ctrtris}, we find
    $$ L^f(z)  :=\d_\si - \ad^\G_{\Ad^D_fj^\F+\d_\si ff^{-1} +\hat O(z)\Ad^D_fj^\perp}  =$$\be =\Ad^G_f \left (\d_\si - \ad^\G_{j^\F+ \hat O(z)j^\perp}\right)\Ad^G_{f^{-1}}  =\Ad^G_f L(z)\Ad^G_{f^{-1}}.\label{adj}\ee
   We thus observe that the spectral invariants
   of the Lax matrix  $L(z)$   are invariant with respect to the action of the loop group $LF$ and therefore they restrict to the well defined functions on the reduced phase space. Moreover, by the general theory of symplectic reduction, the reduced Poisson brackets of the restricted spectral invariants can be evaluated 
   by restricting the unreduced Poisson brackets of the unrestricted invariants.
   In particular, if the unreduced brackets vanish on the constraint surface, the reduced ones  vanish. Said in other words, the  reduced spectral invariants (which are automatically   the integrals of the reduced motion)  Poisson-commute if the unreduced Poisson brackets of the unreduced spectral invariants vanish on the constraint surface $j^{\E\F}=0$. This occurs precisely if the following condition holds
   \be \{L(z)\otimes {\rm Id},{\rm Id}\otimes L(w)\}_{\rm unred}\approx[r(z,w),L(z)\otimes {\rm Id}]-[r^p(w,z),{\rm Id}\otimes L(w)].\label{sid}\ee
   Here $r(z,w)$ is some $r$-matrix acting on the space $L\G\otimes L\G$ and         the symbol $\approx$ indicates that the left-hand-side is equal to the right-hand-side on the constraint surface $j^{\E\F}=0$.
         Said in other words, we first evaluate the unreduced brackets of the Lax operator with itself, then we restrict the result to the constraint surface $j^{\E\F}=0$ and we require that the succession of these two operations gives the right-hand-side of \eqref{sid} also restricted to the constraint surface.
   
  If we choose the   ansatz \eqref{ans1}  
  \be r(z,w)=C_{AB}\ad_{\hat r(z,w)T^A}\otimes\ad_{T^B}\delta(\si_1-\si_2),\label{ans1bis}\ee
  then, similarly as in Section 2.3, we deduce from the conditions \eqref{1e} and \eqref{2e} the validity of the strong integrability condition \eqref{sid}. Indeed, we use    the current Poisson bracket \eqref{spc} to calculate the unreduced Poisson brackets of the matrix elements of the Lax operator
$$\{(O(z)j(\si_1),x)_\G,( O(w)j(\si_2),y)_\G\}_{\rm unred}=$$
$$ =(j(\si_1),[O^\dagger(z)x,O^\dagger(w)y]_\D)_\D\delta(\si_1-\si_2)  +(O^\dagger(z)x,O^\dagger(w)y)_\D\d_{\si_1}\delta(\si_1-\si_2)= $$
  $$= -\Bigl(j(\si_1),O^\dagger(z)[x,\hat r(z,w)y]_\G+ O^\dagger(w)[\hat r(w,z)x,y]_\G +\phi\Bigr)_\D\delta(\si_1-\si_2)$$\be -\Bigl((x,\hat r(z,w)y)_\G+(\hat r(w,z)x,y)_\G\Bigr)\d_{\sigma_1}\delta(\si_1-\si_2) .\label{porbis}\ee
  Here $\phi$ as an unspecified element of $\F\subset\D$ which is eventually  irrevelant because \be (j(\si_1),\phi)_\D=(j^{\E\F}(\si_1),\phi)_\D\approx 0.\ee 
  We conclude the argument by remarking that the desired relation \eqref{sid} is nothing but the operator form of  the equation \eqref{porbis}.

    \section{$\sigma$-models on symmetric spaces}
    \setcounter{equation}{0}
    
    In the present section, we test succesfully our method by applying it  to  three $\sigma$-models on symmetric spaces for which the strong 
    integrability has been already proven in the literature.  
    We first represent all those three models as the dressing cosets and then we 
 readily formulate and solve for them the integrability conditions \eqref{sufdr}, \eqref{1e} and \eqref{2e}. Our unified treatment does not give exactly the same Lax matrices (nor, for that matter, the same $r$-matrices) that the ones   obtained in literature previously, however, our Lax matrices differ from those previously known just outside of the
 primary constraints surface. This means that  they give the same set of the integral of motion in involution as the old Lax matrices do.

       \subsection{The non-deformed symmetric space $\sigma$-model}
    Let $\rho:G\to G$ be an involutive automorphism of  a  quadratic  Lie group $G$ and denote $H$ the subgroup of $G$ consisting of fixed points of this automorphism.  The tangent involutive automorphism $\rho_*:\G\to\G$ has two eigenspaces: to the eigenvalue $+1$ it corresponds the Lie algebra $\H$ and  to the eigenvalue $-1$ it corresponds an ad$_\H$-module $\H^\perp$. As the notation suggests, the eigenspaces $\H$ and $\H^\perp$ are orthogonal to each other with respect to the ad-invariant non-degenerate symmetric bilinear form $(.,.)_\G$ on $\G$.  We note also that the Lie bracket of two elements from $\H^\perp$ belongs to $\H$.   
    
    The space of left cosets $H\backslash G$ is referred to as the
    symmetric space. We now construct the dressing coset which gives rise to an integrable $\si$-model living on the symmetric space target $H\backslash G$. For the Drinfeld double $D$ we take the cotangent bundle $T^*G$  and we view the elements of $D$ as pairs $(g,\gamma)$, $g\in G$, $\ga\in\G$.  The group  multiplication law then reads
     \be (g,x)(m,y)=(gm,x+{\rm Ad}_{g}y), \qquad g,m\in G,\quad x,y\in\G.\ee
     Furthermore, the unit element $e_D$ of $D=T^*G$ is 
      \be e_D=(e_G,0)\label{unitD}\ee 
and the inverse element is
     \be (g,x)^{-1}=(g^{-1},-Ad_{g^{-1}}x).\label{inverseD}\ee
     The involutive automorphism $\rho:G\to G$ can be naturally lifted to the involutive automorphism $\rho_D:D\to D$ by the formula
     \be \rho_D(g,x):=(\rho(g),\rho_*x).\label{rd}\ee
     We denote by $D_H$ the subgroup of the fixed points of the automorphism $\rho_D$. The elements of $D_H$ are obviously the pairs $(h,u)$, $h\in H$, $u\in\H$.
     
     \medskip

The elements of the Lie algebra $\D$ are pairs $(\xi_0,\xi_1)$, $\xi_{i}\in\G$, $i=0,1$,  with the Lie bracket given by
\be [(\xi_0,\xi_1),(\chi_0,\chi_1)]_\D=([\xi_0,\chi_0]_\G,[\xi_0,\chi_1]_\G+[\xi_1,\chi_0]_\G).\label{lb}\ee
The symmetric non-degenerate ad-invariant  bilinear form $(.,.)_\D$ on the Lie algebra $\D$ is given by
   \be \Bigl((\xi_0,\xi_1),(\chi_0,\chi_1)\Bigr)_\D=(\xi_0,\chi_1)_\G+
   (\xi_1,\chi_0)_\G, \ee
   where $(.,.)_\G$ is the symmetric non-degenerate ad-invariant  bilinear form on $\G$. In this paper, $\G$ will always be semi-simple and $(.,.)_\G$ its standard (negatively definite) Killing-Cartan form. 
Note also that the form $(.,.)_\D$ has the split signature 
   $(+,\dots,+,-,\dots,-)$.
    
 \medskip

We shall also need formulas for the left and right Maurer-Cartan forms on $D$
   \be l^{-1}dl\equiv(g,x)^{-1}d(g,x)=(g^{-1}dg,\Ad_{g^{-1}}(dx)),\label{lmc}\ee\be dll^{-1}\equiv d(g,x)(g,x)^{-1}=(dgg^{-1},dx+[x,dgg^{-1}]),\label{rmc}\ee
   and the formula expressing the adjoint action of $D$ on $\D$
   \be \Ad_{(g,x)}(\xi_0,\xi_1)=(\Ad_g \xi_0,\Ad_g \xi_1+\ad_x\Ad_g\xi_0).\label{ada}\ee
   
   \medskip
   
 Now we define the non-degenerate $\E$-model on $D$ by  considering an involution $\E:\D\to\D$ defined as 
   \be \E(\xi_0,\xi_1)=  -(\xi_1,\xi_0).\label{eeta}\ee
   This involution verifies all required properties, it is symmetric with respect to the bilinear form $(.,.)_\D$ and
   the bilinear form $(.,\E.)_\D$ is strictly positive definite. Indeed,
   we have 
      \be \Bigl((\xi_0,\xi_1),\E(\xi_0,\xi_1)\Bigr)_\D=-(\xi_0,\xi_0)_\G-
   (\xi_1,\xi_1)_\G.\ee
   
   \medskip
   
   In order, to define the degenerate $\hat\E$-model we need
   the isotropic subgroup $F\subset D$ such that the adjoint action of $F$ on $\D$ commutes with the involution $\E$. We choose
   for $F$ the group $H$, viewed as the subgroup  of $D$. We can now identify the subspaces featuring in the decomposition \eqref{dirsum} of the Drinfeld double $\D$:
   \be (\phi,0)\in\F,\quad (0,\phi)\in\E\F,\quad (\xi,\mp\xi)\in V_\pm, \qquad \phi\in\H,\quad \xi\in\H^\perp.\ee
 
 In order to obtain the $\si$-model action \eqref{2ndd} corresponding to our degenerate $\hat\E$-model, we have to choose
 the half-dimensional  isotropic subgroup $K\subset D$. We choose
 for $K$ the vector space $\G$ viewed as the Abelian Lie group the elements of which have the form $(e_G,\ga)\in D$. We fix  the $K$-part of the gauge
 symmetry \eqref{gsbis} by setting $l=g\in G$, we then find  
 \be W^\pm_g(\d_\pm gg^{-1},0)=(\d_\pm gg^{-1},\mp P^\perp\d_\pm gg^{-1})\ee
 and, finally, we conclude from \eqref{2ndd}
 \be \label{sys} S_{\hat\E}(g)=  -\frac{1}{2} \int d\tau\oint \biggl( P^\perp\partial_+ gg^{-1},P^\perp\partial_- gg^{-1}\biggr)_\G.\ee
      Here $P^\perp:\G\to \H^\perp$ is the projector with the kernel $\H$. We notice that the action \eqref{sys} has the residual gauge symmetry $g\to hg$, it therefore lives indeed on the symmetric space target $H\backslash G$. 
      
      \medskip
      
     Now we study the integrability of the dressing coset \eqref{sys}. Following the analysis of Section 3.2, we look
     for the families of operators $\hat O(z)$ and $\hat r(z,w)$ verifying the conditions \eqref{sufdr}, \eqref{1e} and \eqref{2e} of
     Section 3.2. We choose the ansatz given by the multiplication by the numerical functions
     \be \hat O(z)(\xi,\mp \xi)=a_\pm(z)\xi,\qquad \xi\in\H^\perp;\label{ho}\ee
           \be \hat r(z,w)\phi=b(z,w)\phi,\quad \hat r(z,w)\xi=c(z,w)\xi, \qquad \phi\in\H,\quad \xi\in\H^\perp.\label{hr}\ee
           We immediately observe that the choice \eqref{ho} respects the condition \eqref{sufdr} provided it holds
           \be a_+(z)a_-(z)=1.\label{apm}\ee
           The dressing coset \eqref{sys} is therefore at least weakly integrable; we may choose without loss of generality
           \be a_\pm(z)=z^{\pm 1}.\label{lpm}\ee
           With the choice \eqref{lpm}, we find the operators $O^\dagger(z)$ 
           \be O^\dagger(z)\phi=(0,\phi), \quad \phi\in\H,\qquad O^\dagger (z)\xi=\frac{1}{2z}(\xi,\xi)-\frac{z}{2}(\xi,-\xi),\quad \xi\in\H^\perp.\label{Odag}\ee
         The conditions \eqref{1e} and \eqref{2e} then give
            \begin{subequations}\label{bc}
        \begin{align}
            b(z,w)+b(w,z)&=0\\
          c(z,w)+c(w,z)&=\jp u(zw)  \\
          b(z,w)z^{\pm 1}+c(w,z)w^{\pm 1}&=\jp u(z),
         \end{align}
              \end{subequations}
              where
              \be u(x):=x-x^{-1}\label{u}.\ee
              The system \eqref{bc} does have  a(n unique) solution  given by
              \be b(z,w)=\jp\frac{u(z)u(w)}{u(w/z)}, \quad  c(z,w)=\jp\frac{u(w)^2}{u(w/z)},\ee
              we thus conclude that the symmetric space dressing coset \eqref{sys} is strongly integrable.
\begin{rem}  {\small         
The strong integrability of the symmetric
space $\si$-model \eqref{sys} was already established in Refs.\cite{DMV12, DMV13}. The Lax matrix of those references is not the same as ours, however, it differs from ours only outside the
constraint surface $j^{\E\F}=0$ (or $\Pi^{(0)}=0$ in the notation of \cite{DMV12,DMV13}). This means
that the reduced spectral invariants of our Lax matrix coincide with those of the Lax matrix of Refs.\cite{DMV12, DMV13}. It may be said also  that our Lax matrix \eqref{Laxfh} is simpler than that of Refs.\cite{DMV12, DMV13} because it does not contain the variable $j^{\E\F}$ but the price to pay for it is that our $r$-matrix
$\hat r(\lm,\eta)$ is more complicated than that of Ref.\cite{DMV12, DMV13} because it is not given by a multiple of identity. }
\end{rem}
   \subsection{Symmetric space $\lm$-deformation   }
     As in the previous section,   let $\rho:G\to G$ be the involutive automorphism of  a  semi-simple  Lie group $G$ and denote $H$ the subgroup of $G$ consisting of fixed points of this automorphism.  However,
     for the double $D$ we now take the direct product $D=G\times G$
             and for the automorphism
          $\rho_D$ we take
          \be \rho_D(a_L,a_R):=(\rho(a_L),\rho(a_R)),\quad (a_L,a_R)\in G\times G.\ee
          The symmetric non-degenerate ad-invariant  bilinear form $(.,.)_\D$ on the Lie algebra $\D$ is now given by
   \be \Bigl((\mu_L,\mu_R),(\nu_L,\nu_R)\Bigr)_\D=(\mu_L,\nu_L)_\G -
   (\mu_R,\nu_R)_\G,\quad \mu_{L,R},\ \nu_{L,R}\in\G,\ee
   where $(.,.)_\G$ is  the (negatively definite) Killing-Cartan form as in Section 4.1. Again,
 the form $(.,.)_\D$ has the split signature 
   $(+,\dots,+,-,\dots,-)$.
             
   \medskip
   
 Now we define the non-degenerate $\E$-model on $D$ by  considering an involution $\E_\al:\D\to\D$ defined as 
   \be \E_\al(\mu_L,\mu_R)=  \cosh{\al}(-\mu_L,\mu_R)+\sinh{\al}(-\mu_R,\mu_L) .\label{ela}\ee
   This involution verifies all required properties, it is symmetric with respect to the bilinear form $(.,.)_\D$ and
   the bilinear form $(.,\E_\al.)_\D$ is strictly positive definite. Indeed,
   we have 
      \be \Bigl((\mu,\nu),\E_\al(\mu,\nu)\Bigr)_\D=-\frac{e^\al}{2}(\mu+\nu,\mu+\nu)_\G-
  \frac{e^{-\al}}{2}(\mu-\nu,\mu-\nu)_\G,\quad \mu,\nu\in\G.\ee
   
   \medskip
   
   In order to define the degenerate $\hat\E$-model we need
   the isotropic subgroup $F\subset D$ such that the adjoint action of $F$ on $\D$ commutes with the involution $\E_\al$. We choose
   for $F$ the subgroup
   \be F=\{(f,f)\in D,\quad f\in H\},\label{fg}\ee
   where, as in Section 4.1,   $H$ is the subgroup of $G$ consisting of the fixed points of the involutive automorphism $\rho:G\to G$.   We can now identify the subspaces featuring in the decomposition \eqref{dirsum} of the Drinfeld double $\D$:
   \be (\phi,\phi)\in\F,\quad (-\phi,\phi)\in\E\F,\quad (\lm^{\pm 1}\xi,\xi)\in V_\pm, \qquad \phi\in\H,\quad \xi\in\H^\perp,\ee
   where
   \be \lm:=\frac{1-e^{\al}}{1+e^{\al}}.\ee
  In order to obtain the $\si$-model action \eqref{2ndd} corresponding to our $\lm$-deformed degenerate $\hat\E$-model, we have to choose
 the half-dimensional  isotropic subgroup $K\subset D$. We choose
 for $K$ the diagonal subgroup of $D$, which means
  \be K=\{(x,x)\in D,\quad x\in G\}.\label{fgbis}\ee
 We fix  the $K$-part of the gauge
 symmetry \eqref{gsbis} by setting $l_0=(g,e_G)\in D$, we then find  
$$W^\pm_g\d_\pm l_0l_0^{-1}= W^\pm_g(\d_\pm gg^{-1},0)=$$\be=-((\lm^{\pm 1}P^\perp +P)\left(\Ad_g-\lm^{\pm 1}P^\perp -P\right)^{-1}\d_\pm gg^{-1},\left(\Ad_g-\lm^{\pm 1}P^\perp -P\right)^{-1}\d_\pm gg^{-1})\ee
   Here $P:\G\to \H$ is the projector with the kernel $\H^\perp$.
 Finally, we conclude from \eqref{2ndd}
 $$ S_\lm(g)=\frac{1}{4} \int d\tau\oint \biggl(\partial_+g g^{-1},  \d_- gg^{-1}  \biggr)_\G+\frac{1}{4}\int \delta ^{-1}\oint \biggl(\delta gg^{-1},[\partial_\sigma gg^{-1},\delta gg^{-1}]\biggr)_\G $$    \be +\frac{1}{2} \int d\tau\oint \biggl(\partial_+g g^{-1},\frac{ 1}{\Ad_g(\lm P^\perp+P)-1} \d_- gg^{-1}  \biggr)_\G\label{sysbis}\ee
     We notice that the action \eqref{sysbis} has the residual gauge symmetry $g\to hgh^{-1}$, $h\in H$. 
     
     \medskip

      Now we study the integrability of the dressing coset \eqref{sysbis}. Following the analysis of Section 3.2, we look
     for the families of operators $\hat O(z)$ and $\hat r(z,w)$ verifying the conditions \eqref{sufdr}, \eqref{1e} and \eqref{2e} of
     Section 3.2. We choose the ansatz given by the multiplication by the numerical functions
     \be \hat O(z)(\lm^{\pm 1}\xi, \xi)=a_\pm(z)\xi,\qquad \xi\in\H^\perp;\label{hobis}\ee
           \be \hat r(z,w)\phi=b(z,w)\phi,\quad \hat r(z,w)\xi=c(z,w)\xi, \qquad \phi\in\H,\quad \xi\in\H^\perp.\label{hrbis}\ee
           We immediately observe that the choice \eqref{hobis} respects  the condition \eqref{sufdr} provided it holds
           \be a_+(z)a_-(z)=1.\label{apmbis}\ee
           The dressing coset \eqref{sysbis} is therefore at least weakly integrable; we may choose without loss of generality
           \be a_\pm(z)=\lm^{\pm\jp} z^{\pm 1}.\label{lpmbis}\ee
           With the choice \eqref{lpmbis}, we find the operators $O^\dagger(z)$ 
           \be O^\dagger(z)(\phi+\xi)=\left(\jp\phi+\frac{u(z\lm^{\jp})}{u(\lm)}\xi,-\jp\phi+\frac{u(z\lm^{-\jp})}{u(\lm)}\xi  \right)   , \quad \phi\in\H, \  \xi\in\H^\perp,\label{Odagbis}\ee
           where
           \be u(x):=x-x^{-1}.\label{ubis}\ee
            The conditions \eqref{1e} and \eqref{2e} then give
         \begin{subequations}\label{bcbis}
        \begin{align}
            b(z,w)+b(w,z)&=0\\
          c(z,w)+c(w,z)&=-\frac{u(zw)}{u(\lm)} \\
          b(z,w)u(z\lm^{\pm\jp})+c(w,z)u(w\lm^{\pm\jp})&=\mp\jp  u(z\lm^{\pm\jp}).
         \end{align}
              \end{subequations}
              The system \eqref{bcbis} does have  a(n unique) solution  given by
              \be b(z,w)=\frac{u(z\lm^{-\jp})u(w\lm^{\jp})}{u(\lm)u(z/w)}-\jp , \quad c(z,w)=\frac{u(w\lm^{-\jp})u(w\lm^{\jp})}{u(\lm)u(z/w)},\ee
              we thus conclude that the $\lm$-deformed symmetric space dressing coset \eqref{sysbis} is strongly integrable.

\begin{rem}  {\small         
The strong integrability of the symmetric
space $\lm$-deformation \eqref{sysbis} was already established in Refs.\cite{HMS15,GSS19}. The Lax matrix of those references is not the same as ours,  but, again, it differs from ours only outside the
constraint surface $j^{\E\F}=0$, which means
that the reduced spectral invariants of our Lax matrix coincide with those coming from the Lax matrix of Refs.\cite{HMS15,GSS19}.  }
\end{rem}

   \subsection{Symmetric space $\eta$-deformation  }

     Let  $\rho:G\to G$ be the involutive automorphism of  the semi-simple Lie group $G$ and denote $H$ the subgroup of $G$ consisting of fixed points of this automorphism. Suppose moroever, that $\rho$ can be lifted 
     to an  involutive automorphism of the complexified  group $\rho_\bc:G^\bc\to G^\bc$. For the Drinfeld double we then take the Lu-Weinstein one \cite{LW} which is the complexified group $D=G^\bc$ viewed  as the real group of doubled dimension. 
          The symmetric non-degenerate ad-invariant  bilinear form $(.,.)_\D$ on the Lie algebra $\D$ is now given by an expression which depends on a real positive parameter $\eta$
   \be (\mu_1+\ri \nu_1,\mu_2+\ri \nu_2)_\D=\frac{1}{\eta}(\mu_1,\nu_2)_\G+ \frac{1}{\eta}(\mu_2,\nu_1)_\G,\quad \mu_{1,2},\nu_{1,2}\in\G.\label{ebil}\ee  Again,
 the form $(.,.)_\D$ turns out to have the split signature 
   $(+,\dots,+,-,\dots,-)$.
             
   \medskip
   
 Now we define the non-degenerate $\E$-model on $D$ by  considering an involution $\E_\eta:\D\to\D$ defined as 
   \be \E_\eta(\mu+\ri\nu)=-\eta^{-1}\nu-\ri\eta\mu,\quad \eta>0, \quad \mu,\nu\in\G.\label{eet}\ee
   This involution verifies all required properties, it is symmetric with respect to the bilinear form $(.,.)_\D$ and
   the bilinear form $(.,\E_\eta.)_\D$ is strictly positive definite. Indeed,
   we have 
      \be  (\mu+\ri\nu,\E_\eta(\mu+\ri\nu)_\D=-(\mu,\mu)_\G-\eta^{-2}(\nu,\nu)_\G,\quad \mu,\nu\in\G.\ee
   
   \medskip
   
   In order to define the degenerate $\hat\E$-model, we choose
   for the isotropic subgroup $F$ the subgroup $H\subset G\subset G^\bc$. The adjoint action of $F$ on $\D=\G^\bc$ then indeed commutes with the involution $\E_\eta$. We can now identify the subspaces featuring in the decomposition \eqref{dirsum} of the Drinfeld double $\D$:
   \be \phi\in\F,\quad \ri\phi\in\E\F,\quad \xi\mp\ri\eta\xi\in V_\pm, \qquad \phi\in\H,\quad \xi\in\H^\perp.\ee
   
  In order to obtain the $\si$-model action \eqref{2ndd} corresponding to our $\eta$-deformed degenerate $\hat\E$-model, we have to choose
 the half-dimensional  isotropic subgroup $K\subset D$. We choose
 for $K$ the subgroup $AN$  of $G^\bc$ featuring in the Iwasava decomposition\footnote{Recall, that $A$ is the non-compact part of the complexified Cartan torus of $G$ and $N$ is the nilpotent subgroup generated by the positive roots. In particular, if $G^\bc$ is $SL(N,\bc)$, then $AN$ consists of uppertriangular matrices with real positive diagonal.} $G^\bc=GAN$. 
 
 In what follows; it will be convenient to parametrize the elements of the Lie algebra $\K$ in terms of those of the Lie algebra $\G$.
 This can be achieved via the $\br$-linear Yang-Baxter operator
 $R:\G^\bc\to\G^\bc$ 
defined as
\be RE^{\al}=-{\rm sign}(\al){\rm i}E^{\al},\quad RH^j=0,\quad [R,\ri]=0,\ee
where $E^\al,H^j$ is the standard  Chevalley basis   of   $\G^{\mathbb C}$ and $\ri$ is viewed as the operator of multiplication by the imaginary unit. Indeed, every element of $\K$ can be then written in an unique way as $(R-\ri)\chi$, where $\chi\in\G$.
 We now fix  the $K$-part of the gauge
 symmetry \eqref{gsbis} by setting $l_0=g\in G$, we then find  
 \be  W^\pm_g\d_\pm l_0l_0^{-1}= W^\pm_g\d_\pm gg^{-1}=\d_\pm gg^{-1}-\eta(R_{g^{-1}}-\ri)(\eta P^\perp R_{g^{-1}} P^\perp\mp 1_{\H^\perp})^{-1}P^\perp\d_\pm gg^{-1},
 \ee
 where \be R_{g^{-1}} =\Ad_g R\Ad_{g^{-1}}.\label{Rg}  \ee
 Finally, we conclude from \eqref{2ndd}
\be S_\eta(g)=-\frac{1}{2} \int d\tau\oint \biggl(P^\perp\partial_+g g^{-1}, \left(1_{\H^\perp}+\eta P^\perp R_{g^{-1}}P^\perp\right)^{-1} P^\perp\d_- gg^{-1}  \biggr)_\G\label{systris}\ee
     We notice that the action \eqref{systris} is the one-parameter deformation of the action \eqref{sys} and it has also the residual gauge symmetry $g\to hg$, $h\in H$. 
     
     \medskip

         Now we study the integrability of the $\eta$-deformed dressing coset \eqref{systris}. Following the analysis of Section 3.2, we look
     for the families of operators $\hat O(z)$ and $\hat r(z,w)$ verifying the conditions  \eqref{sufdr}, \eqref{1e} and \eqref{2e} of
     Section 3.2. We choose the ansatz given by the multiplication by the numerical functions
     \be \hat O(z)\frac{1\mp\ri\eta}{\sqrt{1+\eta^2}}\xi=a_\pm(z)\xi,\qquad \xi\in\H^\perp;\label{hotris}\ee
           \be \hat r(z,w)\phi=b(z,w)\phi,\quad \hat r(z,w)\xi=c(z,w)\xi, \qquad \phi\in\H,\quad \xi\in\H^\perp.\label{hrtris}\ee
           We immediately observe that the choice \eqref{hotris} respects   the condition \eqref{sufdr} provided it holds
           \be a_+(z)a_-(z)=1.\label{apmtris}\ee
           The dressing coset \eqref{systris} is therefore at least weakly integrable; we may choose without loss of generality
           \be a_\pm(z)= z^{\pm 1}.\label{lpmtris}\ee
           With the choice \eqref{lpmtris}, we find the operators $O^\dagger(z)$ 
           \be O^\dagger(z)(\phi+\xi)=\ri\eta\phi +\jp\sqrt{1+\eta^2}\left((1+\ri\eta)z^{-1}-(1-\ri\eta)z\right)\xi, \quad \phi\in\H, \  \xi\in\H^\perp.\label{Odagtris}\ee
       
            The conditions \eqref{1e} and \eqref{2e} then give
         \begin{subequations}\label{bctris}
        \begin{align}
            b(z,w)+b(w,z)&=0\\
          c(z,w)+c(w,z)&=\frac{1+\eta^2}{2}u(zw)\\
        b(z,w)z^{\pm 1}+c(w,z)w^{\pm 1}&=\jp u(z)\mp\frac{\eta^2}{2}s(z)
         \end{align}
              \end{subequations}
                                where
           \be u(x):=x-x^{-1},\quad s(x):=x+x^{-1}.\label{us}\ee
              
              The system \eqref{bctris} does have  a(n unique) solution  given by
              \be b(z,w)=\jp\frac{u(z)u(w)+\eta^2s(z)s(w)}{u(w/z)}, \quad  c(z,w)=\jp\frac{u(w)^2+\eta^2 s(w)^2}{u(w/z)},\ee
              we thus conclude that the $\eta$-deformed symmetric space dressing coset \eqref{systris} is strongly integrable.
    
\begin{rem}  {\small         
The strong integrability of the symmetric
space $\eta$-deformation \eqref{systris} was already established in Ref.\cite{DMV13}. The Lax matrix of those references is not the same as ours,  but, again, it differs from ours only outside the
constraint surface $j^{\E\F}=0$, which means
that the reduced spectral invariants of our Lax matrix coincide with those coming from the Lax matrix of Ref.\cite{DMV13}.  }
\end{rem}

\section{Generalized pseudo-chiral models}
  In all three examples of the integrable dressing cosets treated in Section 4, the
  second term on the right-hand-side of the condition \eqref{sufdr}
 vanished identically. Now we are going to consider examples where it is no longer the case. 
 The simplest one of those new integrable examples  is the pseudo-chiral 
$\sigma$-model of Zakharov and Mikhailov \cite{ZM} characterized by the action
\be S_{\mu}(x)=-\frac{1}{2\mu}\int d\tau \oint \left ((\d_+x,\d_-x)_\G+\frac{\mu}{3}  (x,[\d_+ x,\d_- x]_\G)_\G \right).\label{zm0}\ee 
Here the field $x(\tau,\sigma)$ takes values in the Lie algebra 
$\G$.

\medskip

In the present section, 
we first  interpret the pseudo-chiral model \eqref{zm0} as the dressing coset based on the Drinfeld double\footnote{The interpretation of the $3^{\rm rd}$ order jet bundle 
          of the semi-simple  Lie group $G$ as the Drinfeld double comes from Ref. \cite{LV20}.}  $D=J^3G$, where $J^3G$ is the $3^{\rm rd}$ order jet bundle 
          of the semi-simple  Lie group $G$.
  We then obtain   more complicated integrable dressing cosets  by considering  higher order jet bundles $J^{2n+1}G$. Those theories are apparently new and they describe an integrable interaction of $n$ $\G$-valued fields.
   They represent the principal examples where our method does not just confirm the old results but provides genuinely  new ones.
          
          \subsection{Pseudo-chiral model}
         
         
          Consider the Drinfeld double $D=J^3G$ introduced in \cite{LV20}, where $J^3G$ is the $3^{\rm  rd}$ order jet bundle 
          of the quadratic Lie group $G$. The group $J^3G$ can be conveniently parametrized via the right trivialization
          as the (manifold) direct product $J^3G=G\times \G\times\G\times \G$
          endowed with the group multiplication law\footnote{The multiplication law \eqref{ml} differs from that given in \cite{V,LV20} by  suitable normalization conventions,  namely   $x_i\in\G$ used in \cite{V} are $i!$ multiples of our $x_i$.} and the inverse element
        $$ (g,x_1,x_2,x_3)(h,y_1,y_2,y_3)=$$\be=(gh,x_1+\! ^gy_1,x_2+\! ^gy_2+\jp[x_1,\! ^gy_1],x_3+\! ^gy_3+\frac{2}{3}[x_1,\! ^gy_2]+\frac{1}{3}[x_2,\! ^gy_1]+\frac{1}{6}[x_1[x_1,\! ^gy_1]]),\label{ml}\ee
        \be (g,x_1,x_2,x_3)^{-1}=(g^{-1},-\ ^{g^{-1}}x_1,-\  ^{g^{-1}}x_2, -\ ^{g^{-1}}x_3+\frac{1}{3}\ ^{g^{-1}}[x_1,x_2]),\ee
        where $g,h\in G$, $x_i,y_i\in\G$, $i=1,2,3$ and
        \be \ ^gy_i:=\Ad_g y_i.\label{gx}\ee
        As the vector space, the Lie algebra $\D$ is the direct sum
        $\D=\G\oplus\G\oplus\G\oplus\G$ endowed with the Lie bracket
        {\small \begin{align} [(\xi_0,\xi_1,\xi_2,\xi_3),(\chi_0,\chi_1,\chi_2,\chi_3)]&=\\ = ([\xi_0,\chi_0],[\xi_0,\chi_1]+[\xi_1,\chi_0],[\xi_0,\chi_2]+[\xi_1,\chi_1]+[\xi_2,\chi_0],&[\xi_0,\chi_3]+[\xi_1,\chi_2]+[\xi_2,\chi_1]+[\xi_3,\chi_0]).\end{align}}
        Here, of course, $\xi_i,\chi_i\in\G$, $i=0,1,2,3$.
        
        \medskip

        For convenience, we add the following useful formulas for the exponential map  
        \be \exp{(t(0,\xi_1,\xi_2,\xi_3))}=\left(e_G,t\xi_1,t\xi_2,t\xi_3+\frac{t^2}{6}[\xi_1,\xi_2]\right)\in  J^3G\label{exm}\ee
        and for the right-invariant Maurer-Cartan form on $J^3G$
     \be dll^{-1}
        =\left(dgg^{-1},\Ad_{g}dX_1,\Ad_{g}\left(dX_2+\jp[X_1,dX_1]\right),\Ad_{g}\left(dX_3+[X_1,dX_2]+\frac{1}{6}[X_1,[X_1,dX_1]]\right)\right),\label{mcj3}\ee where  
        now we have parametrized the element $l=(g,x_1,x_2,x_3)$ differently, that is in terms of the products of "pure elements" (cf.\cite{V})
        \be l=(g,0,0,0)(e_G,X_1,0,0)(e_G,0,X_2,0)(e_G,0,0,X_3).\ee
        Note that the different parametrizations of the group $J^3G$ are related by the formulas:
        \be g=g, \quad x_1=\Ad_{g}X_1,\quad x_2=\Ad_gX_2,\quad x_3=\Ad_g\left(X_3+\frac{2}{3}[X_1,X_2]\right).\ee

   The split-signature symmetric non-degenerate ad-invariant  bilinear form $(.,.)_\D$ on the Lie algebra $\D$ is given by
   \be \Bigl((\xi_0,\xi_1,\xi_2,\xi_3),(\chi_0,\chi_1,\chi_2,\chi_3)\Bigr)_\D=(\xi_0,\chi_3)_\G+
   (\xi_1,\chi_2)_\G+(\xi_2,\chi_1)_\G+(\xi_3,\chi_0)_\G.\ee
   
 \medskip
 
 Now we define an appropriate non-degenerate $\E$-model on the
 double $D$ by choosing the following operator
 $\E_\mu:\D\to\D$
 \be \E_\mu(\xi_0,\xi_1,\xi_2,\xi_3)=-(\mu^3\xi_3,\mu\xi_2,\mu^{-1}\xi_1,\mu^{-3}\xi_0).\label{ej3}\ee
 
 To obtain the degenerate $\hat\E$-model, we use the procedure of Section 3.1  and choose the isotropic subalgebra $\F\in\D$ as
 \be \F=\{(\xi_0,0,0,0)\in D,\xi_0\in\G\}.\label{fdef}\ee
 Consequently, the spaces $V_\pm$ are given by
 \be  V_\pm=\{(0,\mu\xi,\mp\xi,0)\in \D,\xi\in\G\}.\label{vpmbis}\ee
 
 We can associate the $\sigma$-model action \eqref{2ndd} to the choices \eqref{ej3}, \eqref{fdef} by choosing
 the maximally isotropic Lie subalgebra $\K\subset \D$ spanned by the elements of $\D$ of the form
 $(0,0,\xi_2,\xi_3)$. Fixing the gauge with respect to both $K$ and $F$ gauge symmetries as $l=(e_G,x,0,0)$ we   find
  \be d(e_G,x,0,0)(e_G,x,0,0)^{-1}=(0,dx,\jp [x,dx],\frac{1}{6}[x,[x,dx]]), \label{mcx}\ee
  therefore the WZ term in the action \eqref{2ndd} is given by the expression
 $$
 \frac{1}{4}\int \delta ^{-1}\oint \biggl(\delta ll^{-1},[\partial_\sigma ll^{-1},\delta ll^{-1}]\biggr)_\D=$$\be=\frac{1}{4}\int \delta ^{-1}\oint (\delta x,[\d_\si x,\delta x]_\G)_\G=\frac{1}{6}\int d\tau\oint (x,[\d_\si x,\d_\tau x]_\G)_\G.\label{wzt}\ee
Using the formula
 \be \Ad_{(e_G,x,0,0)}(0,0,\xi_2,\xi_3)=(0,0,\xi_2,\xi_3+[x,\xi_2]),\ee
 we obtain also 
 \be W_x^\pm\d_\pm ll^{-1} =W_x^\pm \d_\pm(e_G,x,0,0)(e_G,x,0,0)^{-1}=\left(0,\d_{\pm}x,\mp \mu^{-1}\d_{\pm}x, 0 \right),\ee
therefore the dressing coset action \eqref{2ndd} finally reads
\be S_{\mu}(x)=-\frac{1}{2\mu}\int d\tau \oint \left ((\d_+x,\d_-x)_\G+\frac{\mu}{3}  (x,[\d_+ x,\d_- x]_\G)_\G \right).\label{zm}\ee 
We indeed recognize in the formula \eqref{zm} the action of the pseudochiral model \eqref{zm0}.

\medskip

To prove the integrability of the pseudo-chiral model \eqref{zm}, we start with the ansatz
\be \hat O(z)(0,\xi_1,\xi_2,0)=p_1(z)\mu\xi_1+p_2(z)\mu^2\xi_2,\quad \xi_1,\xi_2\in\G,\ee
where $p_1(z)$, $p_2(z)$ are numerical functions that we look for.
The condition \eqref{sufdr} then gives
\be  p_1^2(z)-p_2^2(z)= p_2(z),\ee
which is solved  by
         \be p_1(z)=\frac{- z}{1-z^2}, \quad p_2(z)=\frac{-1}{1-z^2}\label{epc}\ee
        We find also easily the adjoint operator $O^\dagger(z):\G\to\D$ 
        \be O^\dagger(z)\xi= \left(0,p_2(z)\mu^2\xi, p_1(z)\mu\xi,\xi\right).\label{odpc}\ee
        Finally, the operator $\hat r(z,w):\G\to\G$ verifying the conditions \eqref{1e} and \eqref{2e} turns out to be also given by the multiplication by the numerical function
        \be \hat r(z,w)\xi=\frac{\mu^3}{1-w^2}\frac{\xi}{z-w}.\label{pcr}\ee
        In the present context, the polynomial $1-w^2$ is often referred to as the 'twist' function.

 \subsection{ Generalized pseudo-chiral model with two fields}
  
  Now we use our method to construct a genuinely new integrable dressing coset  $\sigma$-model.
  For that we equip the $5^{\rm th}$-order jet bundle $J^{5}G$ with the
  structure of the Drinfeld double.  The group $D=J^{5}G$ can be conveniently parametrized via the right trivialization
          as the direct product of manifolds $J^{5}G=G\times\underbrace{ \G\times\dots\times \G}_{5\ \textrm{times}}$
          endowed with the group multiplication law\footnote{The multiplication law \eqref{inv5} differs from that given in \cite{V,LV20} by  suitable normalization conventions, namely   $x_i\in\G$ used in \cite{V} are $i!$ multiples of our $x_i$.} and the inverse element
        $$ (g,x_1,x_2,x_3,x_4,x_5)(h,y_1,y_2,y_3,y_4,y_5)=$$$$=\Bigl(gh,x_1+\! ^gy_1,x_2+\! ^gy_2+\jp[x_1,\! ^gy_1],x_3+\! ^gy_3+\frac{2}{3}[x_1,\! ^gy_2]+\frac{1}{3}[x_2,\! ^gy_1]+\frac{1}{6}[x_1[x_1,\! ^gy_1]],$$$$x_4+\! ^gy_4+\frac{3}{4}[x_1,\! ^gy_3]+\frac{1}{2}[x_2,\! ^gy_2]+\frac{1}{4}[x_3,\! ^gy_1]+\frac{1}{4}[x_1[x_1,\! ^gy_2]]+\frac{1}{6}[x_2[x_1,\! ^gy_1]]+\frac{1}{12}[x_1[x_2,\! ^gy_1]]+$$$$+\frac{1}{24}[x_1,[x_1,[x_1,\! ^gy_1]]],x_5+\! ^gy_5+ \frac{4}{5}[x_1,\! ^gy_4]+\frac{3}{5}[x_2,\! ^gy_3]+\frac{2}{5}[x_3,\! ^gy_2]+\frac{1}{5}[x_4,\! ^gy_1]+$$$$+\frac{3}{10}[x_1,[x_1,\! ^gy_3]]+\frac{3}{20}[x_3,[x_1,\! ^gy_1]]+\frac{1}{20}[x_1,[x_3,\! ^gy_1]]+\frac{4}{15}[x_2,[x_1,\! ^gy_2]]+\frac{2}{15}[x_1,[x_2,\! ^gy_2]]+$$$$+\frac{1}{10}[x_2,[x_2,\! ^gy_1]]+\frac{1}{15}[x_1,[x_1,[x_1,\! ^gy_2]]]+\frac{1}{20}[x_2,[x_1,[x_1,\! ^gy_1]]]+\frac{1}{30}[x_1,[x_2,[x_1,\! ^gy_1]]]+$$$$+\frac{1}{60}[x_1,[x_1,[x_2,\! ^gy_1]]]+\frac{1}{120}[x_1,[x_1,[x_1,[x_1,\! ^gy_1]]]]\Bigr),\label{ml5}$$
        $$ (g,x_1,x_2,x_3,x_4,x_5)^{-1}=\Bigl(g^{-1},-\ ^{g^{-1}}x_1,-\  ^{g^{-1}}x_2, -\ ^{g^{-1}}x_3+\frac{1}{3}\ ^{g^{-1}}[x_1,x_2],$$$$-\ ^{g^{-1}}x_4+\frac{1}{2}\ ^{g^{-1}}[x_1,x_3]-\frac{1}{12}\ ^{g^{-1}}[x_1,[x_1,x_2]],-\ ^{g^{-1}}x_5+\frac{3}{5}\ ^{g^{-1}}[x_1,x_4]+$$\be  +\frac{1}{5}\ ^{g^{-1}}[x_2,x_3]-\frac{3}{20}\ ^{g^{-1}}[x_1,[x_1,x_3]]+\frac{1}{30}\ ^{g^{-1}}[x_2,[x_2,x_1]]+\frac{1}{60}\ ^{g^{-1}}[x_1,[x_1,[x_1,x_2]]]
        \Bigr) \label{inv5}\ee
        where $g,h\in G$,  $x_i,y_i\in\G$, $i=1,\dots,5$
          and
        \be \ ^gy_i:=\Ad_g y_i.\label{gx5}\ee
        As the vector space, the Lie algebra $\D$ is the direct sum
        $\D=\underbrace{\G\oplus\dots\oplus\G}_{6\ {\rm times}}$ endowed with the Lie bracket
    $$ [(\xi_0,\xi_1,\dots,\xi_5),(\chi_0,\chi_1,\dots,\chi_5)]=([\xi_0,\chi_0],[\xi_0,\chi_1]+[\xi_1,\chi_0],[\xi_0,\chi_2]+[\xi_1,\chi_1]+[\xi_2,\chi_0],$$$$  [\xi_0,\chi_3]+[\xi_1,\chi_2]+[\xi_2,\chi_1]+[\xi_3,\chi_0],  [\xi_0,\chi_4]+[\xi_1,\chi_3]+[\xi_2,\chi_2]+[\xi_3,\chi_1]+[\xi_4,\chi_0],$$\be [\xi_0,\chi_5]+[\xi_1,\chi_4]+[\xi_2,\chi_3]+[\xi_3,\chi_2]+[\xi_4,\chi_1]+[\xi_5,\chi_0]).\ee
        Here, of course, $\xi_i,\chi_i\in\G$, $i=0,1,\dots,5$.
        
        \medskip

        For convenience, we add the following useful formulas for the exponential map  
        \be \exp{(t(0, \dots,0,\xi_j,0,\dots,0))}=(e_G,0, \dots,0,t\xi_j,0,\dots,0)\in  J^{5}G, \quad j=1,...,5\label{exmbis}\ee
    and for the right-invariant Maurer-Cartan form  \be dll^{-1}=d\tilde g\tilde g^{-1}+\Ad_{\tilde g}
     \sum_{j=1}^{5}\Ad_{\prod_{k=1}^{j-1}\tilde X_k}d\tilde X_j\tilde X_j^{-1}, \label{mcj5}\ee where  we parametrized the group
        $J^{5}G$ 
           in terms of the products of the "pure elements"
 $$ l= \tilde g\tilde X_1\tilde X_2\dots \tilde X_{5}\equiv(g,0,0,0,0,0)(e_G,X_1,0,0,0,0)\times $$\be \times(e_G,0,X_2,0,0, 0) (e_G,0,0,X_3,0,0)(e_G,0,0,0,X_4,0) (e_G,0,0,0,0,X_5).\ee
        Note also 
        that 
        \be d\tilde g\tilde g^{-1}=(dgg^{-1},0,0,0,0,0),\ee
        \be d\tilde X_1\tilde X_1^{-1}=
        (0,dX_1,\frac{1}{2!}\ad_{X_1}dX_1,\frac{1}{3!}\ad_{X_1}^2dX_1,\frac{1}{4!}\ad_{X_1}^3dX_1,\frac{1}{5!}\ad_{X_1}^4dX_1),\ee
        \be d\tilde X_2\tilde X_2^{-1}=
        (0,0,dX_2,0,\frac{1}{2!}\ad_{X_2}dX_2,0), \quad  d\tilde X_3\tilde X_3^{-1}=(0,0,0,dX_3,0,0),\ee
          \be d\tilde X_4\tilde X_4^{-1}=
        (0,0,0,0, dX_4,0), \quad  d\tilde X_5\tilde X_5^{-1}=
        (0,0,0,0, 0,dX_5).\label{mcformy}\ee
        
        \medskip
        
        The split-signature symmetric non-degenerate ad-invariant  bilinear form $(.,.)_\D$ on the Lie algebra $\D$ is given by
  \be (\bs\xi,\bs\chi)_\D=\Bigl((\xi_0,\xi_1,\dots,\xi_{5}),(\chi_0,\chi_1,\dots,\chi_{5})\Bigr)_\D=\sum_{j=0}^{5}(\xi_j,\chi_{5-j})_\G.\label{bl5}\ee
   
 \medskip
 
 Now we define an appropriate non-degenerate $\E$-model on the
 double $D$ by choosing the following operator
 $\E_\mu:\D\to\D$
 \be \E_\mu(\xi_0,\xi_1,\xi_2,\xi_3,\xi_4,\xi_5)=-(\mu^{5}\xi_{5},\mu^{3}\xi_{4},\mu\xi_{3},\mu^{-1}\xi_2,\mu^{-3}\xi_1,\mu^{-5}\xi_0).\label{ej5}\ee
 
 To obtain the degenerate $\hat\E$-model, we use the procedure of Section 3.1  choose as the isotropic subalgebra $\F\in\D$ as
 \be \F=\{(\xi_0,0,0,0,0,0)\in D,\xi_0\in\G\}.\label{fdef5}\ee
 Consequently, the spaces $V_\pm$ are formed by the elements of the form 
 \be (0,\mu^3\xi_1,\mu\xi_2,\mp\xi_2,\mp\xi_1,0)\in V_\pm.\label{vpm5}\ee
    We can associate the $\sigma$-model action \eqref{2ndd} to the choices \eqref{ej5}, \eqref{fdef5} by choosing
 the maximally isotropic Lie subalgebra $\K\subset \D$  formed by the elements of the form 
 \be (0,0,0,\xi_3,\xi_4,\xi_5)\in \K.\label{kdef5}\ee
  Fixing the gauge with respect to the both  $K$ and $F$  as $l_0=\tilde X_1\tilde X_2$ we   find from \eqref{mcj5}
  and \eqref{mcformy}$$  d(\tilde X_1\tilde X_2)(\tilde X_1\tilde X_2)^{-1}=
        (0,dX_1,dX_2+\frac{1}{2!}\ad_{X_1}dX_1,\ad_{X_1}dX_2+\frac{1}{3!}\ad_{X_1}^2dX_1,$$\be\frac{1}{2}\ad_{X_2}dX_2+\jp\ad_{X_1}^2dX_2+\frac{1}{4!}\ad_{X_1}^3 dX_1,\jp\ad_{X_1}\ad_{X_2}dX_2+\frac{1}{3!}\ad_{X_1}^3dX_2+\frac{1}{5!}\ad_{X_1}^4dX_1),\label{mc12}\ee
  therefore the WZ term in the action \eqref{2ndd} is given by the expression
 $$
 \frac{1}{4}\int \delta ^{-1}\oint \biggl(\delta l_0l_0^{-1},[\partial_\sigma l_0l_0^{-1},\delta l_0l_0^{-1}]\biggr)_\D= -\frac{1}{5!}\int d\tau\oint  (X_1,[\ad_{X_1}\d_\si X_1,\ad_{X_1}\d_\tau X_1]_\G)_\G+$$$$+\frac{1}{4}\int d\tau\oint (X_2,[\d_\si X_1,\d_\tau X_2]_\G+[\d_\si X_2,\d_\tau X_1]_\G)_\G+$$
 \be \frac{1}{12}\int d\tau\oint \Bigl((\ad_{X_1}\d_\si X_1,\ad_{X_1}\d_\tau X_2)_\G-(\ad_{X_1}\d_\si X_2,\ad_{X_1}\d_\tau X_1)_\G\Bigr).\label{wzt5}\ee 
Using the formula
 \be \Ad_{\tilde X_1\tilde X_2}(0,0,0,\xi_3,\xi_4,\xi_5)=(0,0,0,\xi_3,\xi_4+[X_1,\xi_3],\xi_5+[X_1,\xi_4]+[X_2,\xi_3]+\jp[X_1,[X_1,\xi_3]]),\label{adk5}\ee
 we obtain also 
 $$ W_{X_1,X_2}^\pm\d_\pm l_0l_0^{-1} =W_{X_1,X_2}^\pm \d_\pm (\tilde X_1\tilde X_2)(\tilde X_1\tilde X_2)^{-1}=$$\be =\biggl(0,\d_\pm X_1,\d_\pm X_2+\jp\ad_{X_1}\d_\pm X_1,\mp \mu^{-1}\left(\d_\pm X_2+\jp\ad_{X_1}\d_\pm X_1\right),\mp \mu^{-3}\d_\pm X_1, 0\biggr),\ee
  therefore the dressing coset action \eqref{2ndd} finally reads
$$S_{\mu}(X_1,X_2)=$$$$-\frac{1}{2\mu^3}\int d\tau \oint \left ((\d_+X_1,\d_-X_1)_\G+\mu^2(U_+,U_-)_\G +\mu^3 (X_1,[U_+,U_-]_\G)_\G\right)+$$\be \frac{1}{15}\int d\tau\oint (\ad_{X_1}^3\d_+ X_1, \d_- X_1)_\G   +\frac{1}{6}\int d\tau \oint \left(  (\ad^2_{X_1}\d_+X_1,U_-)_\G-(U_+,\ad^2_{X_1}\d_-X_1)_\G\right),\label{zm5}\ee
  where
 \be U_\pm:=\d_\pm X_2+\jp\ad_{X_1}\d_\pm X_1.\ee
 The action \eqref{zm5} is the 2-field generalization of the pseudo-chiral model of Zakharov et Mikhailov \eqref{zm}. 
 
\medskip

To prove the integrability of the model \eqref{zm5} by using the method developed in Section 3.2, we start with the ansatz
\be \hat O(z)(0,\xi_1,\xi_2,\xi_3,\xi_4,0)=\sum_{j=1}^{4}p_j(z)\mu^j\xi_j.\ee
 The integrability condition \eqref{sufdr} then require that
the four unknown functions $p_j(z)$ are solutions of  
           the following    system of  four equations  
           \be p_1^2(z)-p_4^2(z)=p_2(z),\quad p_2^2(z)-p_3^2(z)=p_4(z)\ee
              \be p_1(z)p_2(z)-p_3(z)p_4(z)=p_3(z),\quad p_1(z)p_3(z)- p_2(z)p_4(z)=p_4(z),\ee
              or, equivalently,
           \be p_i(z)p_j(z)-p_{5-i}(z)p_{5-j}(z)=p_{i+j}(z),\quad 1\leq i\leq j\leq i+j\leq 4.\label{fs5} \ee
           The solution is 
           \be p_1(z)=\frac{z^3-2z}{z^4-3z^2+1},\quad p_2(z)=\frac{z^2-1}{z^4-3z^2+1},\ee \be p_3(z)=\frac{z}{z^4-3z^2+1}, \quad p_4(z)=\frac{1}{z^4-3z^2+1},\quad \ee

        We find also easily the adjoint operator $O^\dagger(z):\G\to\D$
        \be O^\dagger(z)\xi= \left(0,p_4(z)\mu^4\xi,p_{3}(z)\mu^3\xi,p_2(z)\mu^2\xi,p_1(z)\mu\xi,\xi\right). \label{odpc5}\ee
        Finally, we need the operator $\hat r(z,w):\G\to\G$ verifying  the conditions \eqref{1e} and \eqref{2e}. We assume that $\hat r(z,w)$ is given by the  multiplication by a numerical function $\rho(z,w)$, that is
    \be \hat r(z,w)\xi=\mu^5\rho(z,w)\xi.\ee
    This assumption  works, because the conditions \eqref{1e} and \eqref{2e} become
    \be \sum_{j=0}^{m}p_{5-j}(z)p_{5+j-m}(w)+p_{5-m}(z)\rho(z,w)+p_{5-m}(w)\rho(w,z)=0,\label{conm}\ee
    where $ \ m=1,\dots,5$ and we defined $p_0(z)=1$ and $p_{5}(z)=0$.
   The conditions \eqref{conm} are then indeed solved by
        \be \rho(z,w)=\frac{-1}{w^4-3w^2+1}\frac{1}{z-w}\label{pcr5}.\ee
            In the present context, the polynomial $-(w^4-3w^2+1)$ is often referred to as the 'twist' function.
 
 \medskip

 \subsection{$n$-field generalization of the pseudochiral model}
  
  Taking an   integer $n$, we now equip the $(2n+1)^{\rm th}$-order jet bundle $J^{2n+1}G$ with the
  structure of the Drinfeld double.  The group $D=J^{2n+1}G$ can be conveniently parametrized via the right trivialization
          as the direct product of manifolds $J^{2n+1}G=G\times\underbrace{ \G\times\dots\times \G}_{(2n+1)-\textrm{times}}$
          endowed with the group multiplication law\footnote{The multiplication law \eqref{ml} differs from that given in \cite{V,LV20} by  suitable normalization conventions, namely   $x_i\in\G$ used in \cite{V} are $i!$ multiples of our $x_i$.} and the inverse element
        $$ (g,x_1,\dots,x_j,\dots)(h,y_1,\dots y_j,\dots)=$$\be=\Big(gh,x_1+\Ad_g y_1,\dots,x_j+\sum_{i_1+\dots +i_l=j}M_{i_1\dots i_l}\ad_{x_{i_{l-1}}}\dots \ad_{x_{i_{1}}}\Ad_g y_{i_l},\dots\Big),\label{mln}\ee
        $$ (g,x_1,\dots,x_j,\dots)^{-1}=$$\be=(g^{-1},-\Ad_{g^{-1}}x_1,\dots, \sum_{i_1+\dots + i_l=j}(-1)^lM_{i_1\dots i_l}\Ad_{g^{-1}}\ad_{x_{i_1}}\dots \ad_{x_{i_{l-1}}} x_{i_l},\dots),\label{invm}\ee
        where $g,h\in G$,  $x_i,y_i\in\G$, $i=1,\dots,2n+1$ and the numerical coefficients 
        $M_{i_1\dots i_l}$ are given by the formula
        \be M_{i_1\dots i_l}=\prod_{k=1}^l\frac{i_k}{\sum_{m=1}^ki_m}\equiv\frac{i_1}{i_1}\times \frac{i_2}{i_1+i_2}\times\frac{i_3}{i_1+i_2+i_3}\times \dots \times\frac{i_l}{i_1+\dots +i_l}.\ee
        Note also that $i_m$ are strictly positive integers and the sum $\sum_{i_1+\dots i_l=j}$ runs over all possible partitions of the integer $j$. The order of the partition also matters, e.g.
      for the integer $j=3$ we have the partitions $3$, $2+1$, $1+2$ and $1+1+1$  or, in more detail, $i_1=3$, for $l=1$, $i_1=2$, $i_2=1$ as well as $i_1=1$, $i_2=2$ for $l=2$ and $i_1=i_2=i_3=1$ for $l=3$.  The corresponding coefficients $M_{i_1\dots i_l}$ read (cf. \eqref{ml})
      \be M_{3}=1,\quad M_{12}=\frac{2}{3},\quad M_{21}=\frac{1}{3}, \quad M_{111}=\frac{1}{6}.\ee
      We remark in particular that $M_{12}\neq M_{21}$ which is another illustration of the fact that the order of the partition does matter.
        
        \medskip

        As the vector space, the Lie algebra $\D$ is the direct sum
        $\D=\underbrace{\G\oplus\dots \oplus\G}_{(2n+2)-\textrm{times}}$ endowed with the Lie bracket
        \be [\boldsymbol{\xi},\bs\chi]_\D\equiv [(\xi_0,\xi_1,\dots, \xi_{2n+1}),(\chi_0,\chi_1,\dots, \chi_{2n+1})]_\D= ([\bs\xi,\bs\chi]_0,[\bs\xi,\bs\chi]_1,\dots,  [\bs\xi,\bs\chi]_{2n+1}),\label{ncom}\ee 
        where
        \be  [\bs\xi,\bs\chi]_m:= \sum_{j=0}^{m}[\xi_j,\chi_{m-j}]_\G, \quad m=0,\dots,2n+1.\ee
        Here, of course, $\bs\xi,\bs\chi\in\D$, $\xi_i,\chi_i\in\G$, $i=0,1,\dots,2n+1$.
        
        \medskip

        For convenience, we add the following useful formulas for the exponential map  
        \be \exp{(t(0, \dots,0,\xi_j,0,\dots,0))}=(e_G,0, \dots,0,t\xi_j,0,\dots,0)\in  J^{2n+1}G\label{exmtris}\ee
    and for the right-invariant Maurer-Cartan form  \be dll^{-1}=d\tilde g\tilde g^{-1}+\Ad_{\tilde g}
     \sum_{j=1}^{2n+1}\Ad_{\prod_{k=1}^{j-1}\tilde X_k}d\tilde X_j\tilde X_j^{-1}\label{mcjn}\ee where  we parametrized the group
        $J^{2n+1}G$ 
           in terms of the products of the "pure elements"
 $$ l= \tilde g\tilde X_1\tilde X_2\dots\tilde X_{2n+1}\equiv $$\be\equiv(g,0,\dots,0)(e_G,X_1,0,\dots,0)(e_G,0,X_2,0,\dots, 0)\dots (e_G,0,\dots,0,X_{2n+1}).\label{pren}\ee
        Note also 
        that 
        \be d\tilde g\tilde g^{-1}=(dgg^{-1},0,\dots,0),\ee
        \be  d\tilde X_j\tilde X_j^{-1}=(0,\dots, 0, dX_j,0,\dots,0, \underbrace{\jp[X_j,dX_j]}_{(2j)^{\rm th}{\rm place}},0,\dots, 0, \underbrace{\frac{1}{n!}\ad_{X_j}^{n-1}dX_j}_{(nj)^{\rm th}{\rm place}},0,\dots). \label{mcxj}\ee

   The split-signature symmetric non-degenerate ad-invariant  bilinear form $(.,.)_\D$ on the Lie algebra $\D$ is given by
  \be (\bs\xi,\bs\chi)_\D=\Bigl((\xi_0,\xi_1,\dots,\xi_{2n+1}),(\chi_0,\chi_1,\dots,\chi_{2n+1})\Bigr)_\D=\sum_{j=0}^{2n+1}(\xi_j,\chi_{2n+1-j})_\G.\ee
   
 \medskip
 
 Now we define an appropriate non-degenerate $\E$-model on the
 double $D$ by choosing the following operator
 $\E_\mu:\D\to\D$
 $$ \E_\mu(\xi_0,\xi_1,\dots,\xi_{2n},\xi_{2n+1})=$$\be=-(\mu^{2n+1}\xi_{2n+1},\mu^{2n-1}\xi_{2n},\dots,\mu^3\xi_{n+2},\mu\xi_{n+1},\mu^{-1}\xi_n,\mu^{-3}\xi_{n-1},\dots,\mu^{-(2n-1)}\xi_1,\mu^{-(2n+1)}\xi_0).\label{ejn}\ee
 
 To obtain the degenerate $\hat\E$-model, we use the procedure of Section 3.1 and  choose the isotropic subalgebra $\F\in\D$ as
 \be \F=\{(\xi_0,0,\dots,0)\in \D,\xi_0\in\G\}.\label{fdefn}\ee
 Consequently, the spaces $V_\pm$ are given by
   \be V_\pm=(0,\mu^{2n-1}\xi_{1},\dots,\mu^3\xi_{n-1},\mu\xi_{n},\mp\xi_n,\mp\xi_{n-1},\dots,\mp\xi_1,0).\label{vjn}\ee

 We can associate the $\sigma$-model action \eqref{2ndd} to the choices \eqref{ejn}, \eqref{fdefn} by choosing
 the maximally isotropic Lie subalgebra $\K\subset \D$ defined as
  \be \K=\{\bs\xi\in D,\xi_j=0,\ j\leq n\}.\label{kdefn}\ee
 that is $\K$ is spanned by the elements of $\D$ of the form
 $(0,\dots,0,\xi_{n+1},\dots,\xi_{2n+1})$. Fixing the gauge with respect to the both  $K$ and $F$  as $l_0=\tilde X_1\tilde X_2\dots \tilde X_n$ we   find
 \be dl_0l_0^{-1}=
     \sum_{j=1}^{n}\Ad_{\prod_{k=1}^{j-1}\tilde X_k}d\tilde X_j\tilde X_j^{-1}\label{mcjnbis},\ee
  therefore the WZ term in the action \eqref{2ndd} is given by the expression
$$
 \frac{1}{4}\int \delta ^{-1}\oint \biggl(\delta l_0l_0^{-1},[\partial_\sigma l_0l_0^{-1},\delta l_0l_0^{-1}]_\D\biggr)_\D=$$
 \be \frac{1}{4}\int \delta ^{-1}\oint \sum_{k=1}^{2n}\sum_{j=0}^{2n+1-k} \biggl(\left(\delta l_0l_0^{-1}\right)_k,[\left(\d_\sigma l_0l_0^{-1}\right)_{j},\left(\delta l_0l_0^{-1}\right)_{2n+1-k-j}]_\G\biggr)_\G,\label{wztn}\ee
  where $\left(\delta l_0l_0^{-1}\right)_j\in \G$ stand for the $j^{\rm th}$ component of the Maurer-Cartan form $\delta l_0l_0^{-1}\in\D$.
Furthermore, using the formula
 \be \Ad_{\tilde X_1\dots \tilde X_n}\K= \K,\ee
 we obtain also 
 $$ W_X^\pm\d_\pm l_0l_0^{-1}=$$
 \be \biggl(0,\left(\d_\pm l_0l_0^{-1}\right)_1, \dots, \left(\d_\pm l_0l_0^{-1}\right)_n,\mp \mu^{-1}\left(\d_\pm l_0l_0^{-1}\right)_n,\dots, \mp \mu^{-(2n-1)}\left(\d_\pm l_0l_0^{-1}\right)_1, 0\biggr)\label{wn}.\ee

Inserting the formulas \eqref{wn} and \eqref{wztn} 
into the second order dressing coset action \eqref{2ndd} we obtain the $n$-field generalization of the pseudochiral model
$$ S_{\mu}(X_1,\dots,X_n)=-\jp\int d\tau \oint \sum_{j=1}^n\mu^{-(2n+1-2j)} \left(\left(\d_+ l_0l_0^{-1}\right)_j,\left(\d_- l_0l_0^{-1}\right)_j\right)_\G +$$$$+ \frac{1}{4}\int d\tau \oint \sum_{j=1}^n \left(\left(\left(\d_+ l_0l_0^{-1}\right)_j,\left(\d_- l_0l_0^{-1}\right)_{2n+1-j}\right)_\G-\left(\left(\d_+ l_0l_0^{-1}\right)_{2n+1-j}, \left(\d_- l_0l_0^{-1}\right)_j\right)_\G\right)+$$
\be + \frac{1}{4}\int \delta ^{-1}\oint \sum_{k=1}^{2n}\sum_{j=0}^{2n+1-k} \biggl(\left(\delta l_0l_0^{-1}\right)_k,[\left(\d_\sigma l_0l_0^{-1}\right)_{j},\left(\delta l_0l_0^{-1}\right)_{2n+1-k-j}]_\G\biggr)_\G.\label{zmn}\ee 
We remark that the action \eqref{zmn} of the $n$-field pseudochiral model employs just  the components
of the Maurer-Cartan form, which means in fact that it is written in the coordinate-invariant way (the first line correspond to the metric part while the second and third one to the Kalb-Ramond part). Of course, we could express it in the coordinates $X_j$ given by the formulas \eqref{pren} by calculating correspondingly the components of the Maurer-Cartan form from \eqref{mcjnbis}. However, the result of such calculation  is quite cumbersome and, anyway, not very illuminating as it can be observed by looking at the 2-field formula \eqref{zm5}, where we did work out this procedure in detail.
 
\medskip

To prove the integrability of the model \eqref{zmn}, we start with the ansatz
\be \hat O(z)(0,\xi_1,\xi_2,\dots,\xi_{2n-1},\xi_{2n},0)=\sum_{j=1}^{2n}p_j(z)\mu^j\xi_j.\ee
 
The (weak) integrability condition \eqref{sufdr} then require that
the $2n$ unknown functions $p_j(z)$ are solutions of  
           the following    system of   $n^2$ equations  
           \be p_i(z)p_j(z)-p_{2n+1-i}(z)p_{2n+1-j}(z)=p_{i+j}(z),\quad 1\leq i\leq j\leq i+j\leq 2n.\label{fs} \ee
           The system is overdetermined for $n>2$, nevertheless it does possess the needed one-parameter family of solutions. Actually, we find those solutions quite easily taking inspiration from
           the quadratic identities holding for the Fibonacci numbers. The result is
           \be p_{2k}(z)=(-1)^k\frac{\cos{(n-k+1/2)\theta}}{\cos{(n+1/2)\theta}}, \  p_{2k-1}(z)= (-1)^{k}\frac{\sin{(n-k+1)\theta}}{\cos{(n+1/2)\theta}}, \label{soln}\ee
           where $\ k=1,\dots, n$ and
           \be z=2\sin{\frac{\theta}{2}}.\label{zth}\ee

        We find also easily the adjoint operator $O^\dagger(z):\G\to\D$ 
        \be O^\dagger(z)\xi= \left(0,p_{2n}(z)\mu^{2n}\xi,p_{2n-1}(z)\mu^{2n-1}\xi,\dots,p_2(z)\mu^2\xi,p_1(z)\mu\xi,\xi\right). \label{odpcn}\ee
       
        Finally, we need the operator $\hat r(z,w):\G\to\G$ verifying  the (strong) integrability conditions \eqref{1e} and \eqref{2e}. We assume that $\hat r(z,w)$ is given by the  multiplication by a numerical function $\rho(z,w)$, that is
    \be \hat r(z,w)\xi=\mu^{2n+1}\rho(z,w)\xi.\ee
    This assumption  works, because the conditions \eqref{1e} and \eqref{2e} become
    \be \sum_{j=0}^{m}p_{2n+1-j}(z)p_{2n+1+j-m}(w)+p_{2n+1-m}(z)\rho(z,w)+p_{2n+1-m}(w)\rho(w,z)=0,\label{conmbis}\ee
    where $ \ m=0,\dots,2n+1$ and we defined $p_0(z)=1$ and $p_{2n+1}(z)=0$.
   The conditions \eqref{conmbis} are then indeed solved by
        \be \rho(z,w)=\frac{(-1)^n\cos{\frac{\psi}{2}}}{2\cos{(2n+1)\frac{\psi}{2}}}\times
        \frac{1}{\sin{\frac{\psi}{2}}-\sin{\frac{\theta}{2}}},\label{pcrn}\ee
 where
 \be z=2\sin{\frac{\theta}{2}}, \quad w=2\sin{\frac{\psi}{2}}.\label{zwth}\ee 
   In the present context, the 'twist' function is $(-1)^{n+1}\frac{\cos{(2n+1)\frac{\psi}{2}}}{\cos{\frac{\psi}{2}}}$ which is the polynomial function if expressed in the variable $w$.

 \medskip
 
 \noindent \begin{rem}To verify, that the function $\rho(z,w)$ given by the expression \eqref{pcrn} indeed solves the conditions \eqref{conmbis}, it is useful to use the identity
 \be f(\phi)+f(\phi+\alpha)+f(\phi+2\alpha)+\dots +f(\phi+n\alpha)=\frac{\sin{(n+1)\frac{\alpha}{2}}}{\sin{\frac{\alpha}{2}}}f\left(\phi+\frac{n\alpha}{2}\right),\ee
 where $f$ stands either for sin or for cos.
 \end{rem}

\section{Conclusions and outlook} In this paper, we formulated the sufficient conditions for the strong integrability of the dressing cosets. We have also reformulated the non-deformed, $\lm$-deformed and $\eta$-deformed symmetric space cosets as the dressing cosets and we have shown that the integrability of those three theories can be established by solving our sufficient conditions. Finally, we have introduced the new class of dressing cosets based on higher order jet bundles of quadratic Lie groups playing the role of the Drinfeld doubles. Those new theories can be interpreted as the (interacting)
$n$-field generalizations of the pseudo-chiral $\sigma$-model of Zakharov and Mikhailov.  We have solved our sufficient conditions of the strong integrability also in this case and identified the so called twist functions of those theories which turn out to have poles only in the infinity.

As far as the outlook is concerned, we 
find appealing to study in future how  our  results could be put in the perspective of two other frameworks which are currently used along with the $\E$-models to study integrable $\sigma$-models on group manifolds, namely, the 
formalism based on the 4d Chern-Simons gauge theory \cite{CY} as well as the one based on the structure of the affine Gaudin models \cite{Vi,DLMV}.

                    \end{document}